	\documentclass{article}

	\usepackage{geometry}
	\usepackage{amsmath}
	\usepackage{amssymb}
	
	\usepackage[hyphens]{url}

	\usepackage{hyperref}
	
	\usepackage[hyphenbreaks]{breakurl}

	\usepackage{empheq}

	\allowdisplaybreaks

	\newcommand{\txt}[1]{\texttt{#1}}

	\usepackage[parfill]{parskip}
	
	\parskip = \baselineskip

	\usepackage{tikz}
	\usepackage{pgfplots}
	\pgfplotsset{compat=newest}

	\usepackage{subcaption}
	
	\usepackage[labelfont=bf]{caption}

	\usepackage{enumitem}

	\usepackage{float}
	
	\floatstyle{plain}
	\newfloat{algorithm}{tbp}{alg}
	\floatname{algorithm}{Algorithm}
	
	\usepackage{caption}
	\usepackage{tabularx}
	
	\usepackage{pseudo}
	
	\pseudodefinestyle{fullwidth}{
		begin-tabular =
		\tabularx{0.9\linewidth}{@{}
			r
			>{\pseudosetup}
			X
			>{\leavevmode\small\color{black!60}}
			p{0.25\linewidth}
			@{}},
		end-tabular = \endtabularx,
		setup-append = \pseudoeq
	}

	\usepackage{booktabs}
	
	\usepackage{calligra}
	\DeclareMathAlphabet{\mathcalligra}{T1}{calligra}{m}{n}
	\DeclareFontShape{T1}{calligra}{m}{n}{<->s*[2.2]callig15}{}
	\newcommand{\scriptr}{\mathcalligra{r}\,}

	\usepackage{stmaryrd}
	\newenvironment{envforbits}{\ensuremath{\llbracket}}{\ensuremath{\rrbracket}}
	\newcommand{\bit}[1]{\begin{envforbits}{#1}\end{envforbits}}

	\def\O{\mathcal{O}}
	
	\def\a{\alpha}
	\def\b{\beta}
	\def\bnz{\eqref{bnz}}
	\def\bs{\eqref{bs}}
	\def\bz{\eqref{bz}}
	\def\cw{\eqref{cw}}
	\def\d{\delta}
	\def\eps{\varepsilon}
	\def\g{\gamma}
	\def\G{\Gamma}
	\newcommand{\I}[2]{I_{#1}[ {#2} ]}
	\def\k{\kappa}
	\def\s{\sigma}
	\def\th{\theta}
	\def\z{\zeta}
	
	\def\av{\vec{\a}}
	\def\bv{\vec{\b}}
	\def\Cv{\vec{C}}
	\def\dv{\vec{\d}}
	\def\Dv{\vec{D}}
	\def\gv{\vec{\g}}
	\def\qv{\vec{q}}
	
	\def\F{\vec{F}}
		
	\def\abspos{\scriptr}
	\def\absposv{\vec{\abspos}}
	\def\absposvprime{ \absposv^{ \, \prime } }
	
	\def\relpos{r}
	\def\relposv{\vec{\relpos}}
	
	\def\w{\vec{w}}
	\def\rhov{\vec{\rho}}
	\def\rhovprime{ \rhov^{ \, \, \prime } }
	\def\rhovprimeprime{ \rhov^{ \, \, \prime \prime } }
	\newcommand{\rhovsup}[1]{ \rhov^{ \, \, (#1) } }
	
	\def\p{\vec{p}}
	
	\def\vwind{\vec{v}^{\,\text{wind}}}
	\def\vrel{\vec{v}^{\, \text{rel}}}
	
	\newcommand{\bq}{\begin{equation}}
	\newcommand{\eq}{\end{equation}}
	
	\def\ourstext{ours}
	\def\numtext{num}

	\newcommand{\subtext}[1]{_{ \textnormal{#1} }}
	\newcommand{\suptext}[1]{^{ \text{#1} }}
	
	\def\air{\subtext{air}}
	\def\ascent{\subtext{ascent}}
	\def\atm{\subtext{atm}}
	\def\c{\subtext{c}}
	\def\comp{\subtext{expanded}}
	\def\decay{\subtext{d}}
	
	\def\drag{\subtext{drag}}
	\def\E{\subtext{E}}
	\def\eff{\subtext{eff}}
	\def\f{\subtext{f}}
	\def\final{\subtext{final}}
	\def\flight{\subtext{flight}}
	\def\grav{\subtext{grav}}
	\def\lin{^{\, \text{lin} }}
	\def\net{\subtext{net}}
	\def\posttan{\suptext{post-tan}}
	\def\pretan{\suptext{pre-tan}}
	\def\safe{\suptext{safe}}
	\def\subarg{\subtext{arg}}
	\def\subext{\subtext{extreme}}
	\def\submax{\subtext{max}}
	
	\def\subnum{\subtext{\numtext}}
	\def\subODE{\subtext{ODE}}
	\def\subours{\subtext{\ourstext}}
	\def\tang{\subtext{tangent}}

	\def\td{\tau\decay}
	\def\exppt{e^{ \tau / \td }}
	\def\expmt{e^{ -\tau / \td }}
	
	\definecolor{mygray}{gray}{0.95}
	\newcommand{\mybox}[1]{\colorbox{mygray}{#1}}

	\newcommand{\addMatlabQuadFigureSmall}[3]{%
		\begin{figure}[H]
			\centering
			\begin{tikzpicture}[scale=1]
				\begin{axis}% "axis equal" and "axis equal image" may help with scaling
					[ width = 0.245\linewidth , name = axis_1 , grid = major , grid style = { dashed , gray!20 } , xlabel = $t$ , ylabel = $\abspos - R\E$ , tick scale binop = \times , scaled y ticks = true , scaled x ticks = true , legend cell align = { left } , legend style = { at = {(3.1,-0.75)} , anchor = center , font = \footnotesize , /tikz/column 2/.style = { column sep = 5pt } , /tikz/column 4/.style = { column sep = 5pt } } , legend columns = 3 , label style = { font = \footnotesize } , tick label style = { font = \footnotesize } , title = \footnotesize{\textbf{Radial Trajectories}} ]
					\addplot
						[ very thick , gray!50 ]
						table[ x = t_s , y = traj_num , col sep = comma ] {#1};
						\addlegendentry{Numerical}
					\addplot
						[ very thick , dashed , blue ]
						table[ x = t_0 , y = traj_0 , col sep = comma ] {#1};
						\addlegendentry{Order 0}
					\addplot
						[ very thick , dotted , red ]
						table[ x = t_1 , y = traj_1 , col sep = comma ] {#1};
						\addlegendentry{Expanded}
				\end{axis}
				
				\begin{semilogyaxis}% "axis equal" and "axis equal image" may help with scaling
					[ width = 0.245\linewidth , name = axis_2 , at = { (axis_1.south east) } , xshift = 0.11\linewidth , grid = major , grid style = { dashed , gray!20 } , xlabel = $t$ , ylabel = $E\subODE$ , legend style = { at = {(1,1)} , anchor = north east } , tick scale binop = \times , scaled y ticks = true , scaled x ticks = true , label style = { font = \footnotesize } , tick label style = { font = \footnotesize } , title = \footnotesize{\textbf{ODE Errors}} ]
					\addplot
						[ very thick , gray!50 ]
						table[ x = t_s , y = ode_err_num , col sep = comma ] {#1};
					\addplot
						[ very thick , dashed , blue ]
						table[ x = t_0 , y = ode_err_0 , col sep = comma ] {#1};
					\addplot
						[ very thick , dotted , red ]
						table[ x = t_1 , y = ode_err_1 , col sep = comma ] {#1};
				\end{semilogyaxis}
				
				\begin{semilogyaxis}% "axis equal" and "axis equal image" may help with scaling
					[ width = 0.245\linewidth , name = axis_3 , at = { (axis_2.south east) } , xshift = 0.11\linewidth , grid = major , grid style = { dashed , gray!20 } , xlabel = $t$ , ylabel = $\Delta_{ \absposv }$ , legend style = { at = {(1,1)} , anchor = north east } , tick scale binop = \times , scaled y ticks = true , scaled x ticks = true , minor tick style = { draw = none } , label style = { font = \footnotesize } , tick label style = { font = \footnotesize } , title = \footnotesize{\textbf{Position Deviations}} ]
					\addplot
						[ very thick , dashed , blue ]
						table[ x = t_0 , y = pos_dev_0 , col sep = comma ] {#1};
					\addplot
						[ very thick , dotted , red ]
						table[ x = t_1 , y = pos_dev_1 , col sep = comma ] {#1};
				\end{semilogyaxis}
				
				\begin{semilogyaxis}% "axis equal" and "axis equal image" may help with scaling
					[ width = 0.245\linewidth , at = { (axis_3.north east) } , anchor = north west , xshift = 0.11\linewidth , grid = major , grid style = { dashed , gray!20 } , xlabel = $t$ , ylabel = $\Delta_{ \absposvprime }$ , legend style = { at = {(1,1)} , anchor = north east } , tick scale binop = \times , scaled y ticks = true , scaled x ticks = true , label style = { font = \footnotesize } , tick label style = { font = \footnotesize } , title = \footnotesize{\textbf{Velocity Deviations}} ]
					\addplot
						[ very thick , dashed , blue ]
						table[ x = t_0 , y = vel_dev_0 , col sep = comma ] {#1};
					\addplot
						[ very thick , dotted , red ]
						table[ x = t_1 , y = vel_dev_1 , col sep = comma ] {#1};
				\end{semilogyaxis}
			\end{tikzpicture}
			\caption{#2}
			\label{#3}
		\end{figure}
	}

\title{A More General Linear Projectile Problem}

\author{Nick Lorenzo\\\footnotesize{\texttt{Lorenzo.Nick@gmail.com}}}

\begin{document}

\maketitle

\begin{abstract}\noindent
In a full 3D context, we study a projectile subject to linear drag, a non-uniform gravitational field, time-dependent wind, and parameterized atmospheric thinning.  In this general context, we provide integral solutions, exact to $\O( \eps )$, for the position and velocity of the projectile, where $\eps$ is a small perturbation parameter; in the special case of constant wind, we provide closed-form solutions, exact to $\O( \eps )$.  Under the constant-wind assumption, we provide closed-form solutions of $\O( 1 )$ for the time of tangency, times of flight, and extreme values of the radius achieved by the projectile.  We provide physical interpretations throughout, including a physical interpretation of the branches $W_0$ and $W_{ -1 }$ of the Lambert W function in the context of flight time.  We also provide parameterized, error-controlled algorithms to compute trajectories, complete with a full Matlab implementation that we make freely available.  We compare the results of our implementation to a general-purpose, stiff ODE solver.
\end{abstract}

\tableofcontents

\section{Introduction}

The linear projectile problem is a classic problem in classical mechanics, receiving an elementary treatment in texts such as \cite[Sec.~2.2]{Taylor}.  More advanced treatments of the problem can be found in works such as \cite{wipm} and \cite{review-wipm}, which study the problem with constant wind.

\subsection{Generalizations investigated in this work}

In this paper, we generalize the typical approach to the linear projectile problem in several ways.

\begin{enumerate}[label={(\textbf{\arabic*})}]

	\item We include the effects of a non-uniform gravitational field.
	
	\item We permit an arbitrary, vector-valued, time-dependent wind function in 3D space.
	
	\item We permit a parameterized atmospheric thinning function.
	
	\item We permit an arbitrary initial position in 3D space, on or above the surface of the Earth.
	
	\item We study the flight time of the projectile with respect to an arbitrary final radius, on or above the surface of the Earth.
	
	\item We study the extreme value of the radius achieved by the projectile, whether that extreme value is a maximum or a minimum.
	
	\item We permit an initial velocity of arbitrary magnitude and direction in 3D space.

\end{enumerate}

\subsection{Contributions of this work}

Under the above generalizations, we provide the following contributions to the literature.

\begin{enumerate}[label={(\textbf{\arabic*})}]

	\item We develop the full 3D, nonlinear, coupled, second-order system of ODEs modeling the problem (Section \ref{sec:model_development}).
	
	\item We provide physical interpretations of the ODEs of both $\O( 1 )$ and $\O( \eps )$, where $\eps$ is a small perturbation parameter (Section \ref{sec:integral_solns}).
	
	\item We provide integral solutions exact to $\O( \eps )$ for the position and velocity of the projectile (Section \ref{sec:integral_solns}).
	
	\item We provide, in the special case of constant wind, closed-form solutions of both $\O( 1 )$ and $\O( \eps )$ for the position and velocity of the projectile (Section \ref{sec:const_wind}), and we present numerical evidence that our $\O( \eps )$ solutions increase the accuracy of our $\O( 1 )$ solutions with respect to the original model (Section \ref{sec:numerical}).
	
	\item We provide an analysis of the validity of our approximations and solutions (Section \ref{sec:model_parameters_and_validity}).
	
	\item We provide general $\O( 1 )$ conditions on the time of tangency, times of flight, and extreme values of the radius achieved by the projectile (Section \ref{sec:const_wind_analysis}).
	
	\item We provide, in the special case of constant wind, closed-form solutions of $\O( 1 )$ for the time of tangency, times of flight, and extreme values of the radius achieved by the projectile, with physical interpretations (Section \ref{sec:const_wind_analysis}).
	
	\item We provide a physical interpretation of the relevant branches of the Lambert W function in the context of flight time (Section \ref{sec:Lambert_discussion}).
	
	\item We provide a parameterized method of controlling the error in our solutions (Section \ref{sec:algorithms}).
	
	\item We provide a full numerical implementation of our solutions with parameterized error control in the constant-wind case, including code to reproduce all the data and figures we present (see \cite{matlab-code}).

\end{enumerate}

\section{Development of the projectile-motion model}\label{sec:model_development}

Consider a non-relativistic, point-like projectile subject only to the force of Earth's gravitational field, $\F\grav$, and to the force of air resistance, $\F\air$.  The net force on the projectile may be written as
\bq
\label{net_force}
\F\net = \F\grav + \F\air.
\eq
Let $m$ be the mass of the projectile, let $g \approx 9.81$ meters / second$^2$ be the acceleration of gravity at the surface of the Earth, let $R\E \approx 6.361 \times 10^6$ meters be the radius of the Earth (which we assume to be spherical and of uniform density), and let $\absposv( t )$ be the position of the projectile at time $t$, with the origin at the center of the Earth.  Then we may write the force of gravity as
\bq
\F\grav = -\frac{ m g R\E^2 }{ \abspos^3 } \absposv,
\eq
where $\abspos( t ) \equiv | \absposv( t ) | \ge R\E$ is the distance from the projectile to the center of the Earth at time $t$.

We assume the force of air resistance to be well-approximated by% \cite[p.~44]{Taylor}
\bq
\F\air \sim \F\air\lin,
\eq
where
\bq
\F\air\lin = -f\atm( s ) b\drag \vrel
\eq
is the force of linear resistance, with linear drag coefficient $b\drag \ge 0$ at sea level.  The function $f\atm : [ 0 , \infty ) \to [ 0 , 1 ]$ is a dimensionless atmospheric thinning function with dimensionless argument $s : [ 0 , \infty ) \to [ 0 , \infty )$ defined by
\bq
s( t ) \equiv \frac{ \abspos( t ) - R\E }{ \ell };
\eq
$f\atm( s( t ) )$ models the density of air at radius $\abspos( t )$ relative to the density of air at radius $R\E$ (sea level).  Here, $\ell > 0$ is a characteristic length scale for the atmospheric thinning function $f\atm$ (see Section \ref{sec:traj_plots} for an example of $f\atm$ and $\ell$).  The quantity
\bq
\vrel \equiv \frac{ d\absposv }{ dt } - \vwind
\eq
is the relative velocity of the projectile through the air, with the motion of the air itself modeled by the vector-valued, time-dependent wind function $\vwind$.

Applying Newton's second law\footnote{Here we neglect the fact that the Earth is a non-inertial, rotating reference frame.  This can be handled by relating our results to a time-dependent coordinate system, but we do not develop that model here.} to the projectile subject to the net force \eqref{net_force}, we find the projectile's motion to be modeled by the nonlinear, coupled, second-order system of ODEs given by
\begin{empheq}[box=\mybox]{equation}
\label{original_model}
m \frac{ d^2 \absposv }{ dt^2 } = -\frac{ m g R\E^2 }{ \abspos^3 } \absposv - f\atm\left( \frac{ \abspos - R\E }{ \ell } \right) \left[ \frac{ d \absposv }{ dt } - \vwind \right] b\drag,
\end{empheq}
with initial conditions
\begin{subequations}
\label{original_initial_conditions}
\begin{empheq}[box=\mybox]{align}
\absposv( t = 0 ) & = \absposv_0 \equiv R_0 \av,\\
\frac{ d\absposv }{ dt }( t = 0 ) & = v_0 \bv.
\end{empheq}
\end{subequations}
Here, $R_0 \equiv \abspos( t = 0 ) \ge R\E$ is the initial distance from the projectile to the center of the Earth, $v_0$ is the initial speed of the projectile, and $\av$ and $\bv$ are vectors given in Cartesian coordinates by
\begin{align}
\av & \equiv ( \sin( \th_r ) \cos( \phi_r ) , \sin( \th_r ) \sin( \phi_r ) , \cos( \th_r ) ),\\
\bv & \equiv ( \sin( \th_v ) \cos( \phi_v ) , \sin( \th_v ) \sin( \phi_v ) , \cos( \th_v ) ) \bit{ v_0 \neq 0 },
\end{align}
where $\th_r \in [ 0 , \pi ]$ is the inclination angle and $\phi_r \in [ 0 , 2 \pi )$ is the azimuthal angle of the initial position of the projectile, and where $\th_v \in [ 0 , \pi ]$ is the inclination angle and $\phi_v \in [ 0 , 2 \pi )$ is the azimuthal angle of the initial velocity of the projectile.  Here, the notation $\bit{ X }$ is the Iverson bracket, evaluated as $1$ if $X$ is true and $0$ otherwise.

\section{Transformation of the model to dimensionless form}\label{sec:model_tfn}% See pg 106 of notes.

Define the relative position $\relposv( t )$ of the projectile by
\bq
\label{relpos_defn}
\relposv( t ) \equiv \absposv( t ) - \absposv_0.
\eq
Let $0 < t\c$ be a characteristic time for the problem, let $0 < r\c$ be a characteristic length for the problem (each to be specified in Section \ref{sec:parameters}), and define the characteristic speed and scalar acceleration
\begin{subequations}
\label{vc_ac}
\begin{align}
v\c & \equiv \frac{ r\c }{ t\c },\\
a\c & \equiv \frac{ r\c }{ t\c^2 }.
\end{align}
\end{subequations}
We then define the dimensionless time $\tau$, relative position $\rhov$, and wind $\w$ by
\begin{subequations}
\label{scaling_eqn}
\begin{align}
\label{tau_eqn}
\tau 			& \equiv \frac{ 1 }{ t\c } t,\\
\rhov( \tau )	& \equiv \frac{ 1 }{ r\c } \relposv( t( \tau ) ),\\
\w( \tau ) 		& \equiv \frac{ 1 }{ v\c } \vwind( t( \tau ) ).
\end{align}
\end{subequations}
Here, from the relationship \eqref{tau_eqn},
\bq
\label{t_t_c_eqn}
t( \tau ) \equiv t\c \tau.
\eq
We find that
\begin{subequations}
\begin{align}
\frac{ d }{ dt } 								& = \frac{ 1 }{ t\c } \frac{ d }{ d\tau },\\
\frac{ d^n \absposv }{ dt^n }					& = \frac{ d^n \relposv }{ dt^n },\\
\frac{ d^n }{ dt^n } \relposv( t( \tau ) )	& = \frac{ r\c }{ t\c^n } \frac{ d^n }{ d\tau^n } \rhov( \tau ).
\end{align}
\end{subequations}
Now, assume $r\c$ is defined in such a way that
\bq
r\c \ll R_0
\eq
is satisfied, and define the small, dimensionless parameter
\bq
\label{eps_defn}
\eps \equiv \frac{ r\c }{ R_0 } \ll 1.
\eq
Also define
\begin{subequations}
\label{eta_u_defn}
\begin{align}
\rho	& \equiv | \rhov |,\\
\eta	& \equiv 2 \eps ( \av \cdot \rhov ) + \eps^2 \rho^2,\\
u 		& \equiv 1 + \eta.
\end{align}
\end{subequations}
Now,
\begin{subequations}
\label{distance_expansion}
\begin{align}
\abspos^k
	& = | \absposv |^k\\
	& = | \absposv_0 + \relposv |^k\\
	& = \left[ | \absposv_0 |^2 + 2 ( \absposv_0 \cdot \relposv ) + | \relposv |^2 \right]^{ k / 2 }\\
	& = \left[ R_0^2 + 2 R_0 ( \av \cdot \relposv ) + | \relposv |^2 \right]^{ k / 2 }\\
	& = \left[ R_0^2 \left( 1 + \frac{ 2 }{ R_0 } ( \av \cdot \relposv ) + \frac{ 1 }{ R_0^2 } | \relposv |^2 \right) \right]^{ k / 2 }\\
	& = R_0^k \left[ 1 + \frac{ 2 }{ R_0 } ( \av \cdot \relposv ) + \frac{ 1 }{ R_0^2 } | \relposv |^2 \right]^{ k / 2 }\\
	& = R_0^k \left[ 1 + 2 \frac{ r\c }{ R_0 }  ( \av \cdot \rhov ) + \frac{ r\c^2 }{ R_0^2 } \rho^2 \right]^{ k / 2 }\\
	& = R_0^k \left[ 1 + 2 \eps ( \av \cdot \rhov ) + \eps^2 \rho^2 \right]^{ k / 2 }\\
 	\label{distance_expansion_u}
 	& = R_0^k u^{ k / 2 }\\
 	& = R_0^k ( 1 + \eta )^{ k / 2 }\\
	\label{distance_expansion_eta}
 	& = R_0^k \left[ 1 + \frac{ 1 }{ 2 } k \eta + \frac{ 1 }{ 8 } k ( k - 2 ) \eta^2 + \O( \eta^3 ) \right]\\
 	\label{distance_expansion_eps}
 	& = R_0^k \left[  1 + k \eps ( \av \cdot \rhov ) + \frac{ 1 }{ 2 } k \eps^2 \{ \rho^2 + ( k - 2 ) ( \av \cdot \rhov )^2 \} + \O( \eps^3 ) \right].
\end{align}
\end{subequations}
The accuracy of the expansion in powers of $\eta$ depends on the value of $\eta$; we later specify this accuracy by placing a bound on $\eta$.  This leads to a bound for $\eps$, and therefore for $r\c$ (see Section \ref{sec:parameters}).  For now, we assume that $\eta \ll 1$.

We note that
\bq
\label{s_u}
s = \k ( u^{ 1 / 2 } - \G ),
\eq
where
\bq
\label{Gamma_defn}
\G \equiv \frac{ R\E }{ R_0 } \le 1
\eq
and
\bq
\k \equiv \frac{ R_0 }{ \ell }.
\eq
With \eqref{Gamma_defn}, \eqref{vc_ac} -- \eqref{eta_u_defn}, \eqref{distance_expansion_u}, and \eqref{s_u} in mind, we find that the original model \eqref{original_model} becomes
\bq
m a\c \frac{ d^2 \rhov }{ d\tau^2 } = -\frac{ m g \G^2 }{ R_0 u^{ 3 / 2 } } ( R_0 \av + r\c \rhov ) - f\atm\left( \k ( u^{ 1 / 2 } - \G ) \right) v\c b\drag \left[ \frac{ d\rhov }{ d\tau } - \w \right],
\eq
which we rewrite more compactly as
\begin{empheq}[box=\mybox]{equation}
\label{dimensionless_model}
\rhovprimeprime = ( \gv + \eps \g \rhov ) u^{ -3 / 2 } - f\atm( s ) b ( \rhovprime - \w ).
\end{empheq}
This is the general form of our dimensionless model, where
\begin{subequations}
\begin{align}
\label{gamma_defn}
\g  & \equiv -\frac{ g \G^2 }{ a\c },\\
\gv & \equiv \g \av,\\
b   & \equiv \frac{ t\c }{ m } b\drag.
\end{align}
\end{subequations}
\paragraph{Physical interpretation} In \eqref{gamma_defn}, the factor $\G^2$ adjusts the value of $g$ to account for values of $R_0$ not equal to $R\E$, while division by the characteristic (scalar) acceleration $a\c$ makes $\g$ dimensionless; we therefore think of $| \g |$ as the dimensionless value of $g$ we would find at a radius $R_0$ from the center of the Earth, and we think of $\gv$ as the corresponding dimensionless, vector-valued acceleration having fixed direction along the $\av$-axis defined by $\{ x \av \mid x \in \mathbb{ R } \}$.  The quantity $b$ is the dimensionless linear drag coefficient.

We note that \eqref{dimensionless_model} is still an exact representation of the original model \eqref{original_model}, as no approximations have yet been made.  Define the dimensionless initial speed and dimensionless initial velocity
\begin{subequations}
\begin{align}
\d	& \equiv \frac{ v_0 }{ v\c },\\
\dv	& \equiv \d \bv.
\end{align}
\end{subequations}
Then we also find the dimensionless version of the initial conditions \eqref{original_initial_conditions} to be given by
\begin{subequations}
\label{nondim_initial_conditions}
\begin{empheq}[box=\mybox]{align}
\rhov( 0 ) & = \vec{ 0 },\\
\rhovprime( 0 ) & = \dv.
\end{empheq}
\end{subequations}
We note that a solution $\rhov\,$ to the IVP defined by \eqref{dimensionless_model} and \eqref{nondim_initial_conditions} necessarily depends on the choice of atmospheric thinning function $f\atm$, its parameter $\ell$, the choice of $r\c$ defining $\eps$, the choice of $t\c$ relating $t$ and $\tau$, the value of the air resistance parameter $b\drag$, and the choice of wind function $\vwind$.

\section{Spatial expansion of the dimensionless model}\label{sec:spatial_expn}

Defining $s_0 \equiv s( t = 0 )$ and using \eqref{distance_expansion}, we find that
\begin{subequations}
\begin{align}
s - s_0
	& = \frac{ \abspos - R_0 }{ \ell }\\
	& = \frac{ 1 }{ \ell } [ r_c ( \av \cdot \rhov ) + \O( \eps ) ]\\
	& = \k \eps ( \av \cdot \rhov ) + \O( \eps^2 ).
\end{align}
\end{subequations}
We assume that $f\atm( s )$ may be written as
\begin{subequations}
\label{f_atm_expansion}
\begin{align}
f\atm( s )
	& = f\atm( s_0 ) + ( s - s_0 ) f\atm'( s_0 ) + \tfrac{ 1 }{ 2 } ( s - s_0 )^2 f\atm''( s_0 ) + \ldots\\
	& = f\atm( s_0 ) + \k \eps ( \av \cdot \rhov) f\atm'( s_0 ) + \O( \eps^2 ).
\end{align}
\end{subequations}
We note that the atmospheric expansion \eqref{f_atm_expansion} is a good approximation when $\k \eps ( \av \cdot \rhov ) \ll 1$.  Due to our later bounding of $\rho$ by $\rho\submax$ (see Section \ref{sec:parameters}), this is guaranteed if
\bq
\label{another_r_c_condn}
\k \eps \rho\submax = \frac{ r\c \rho\submax }{ \ell } \ll 1;
\eq
we assume that $r\c$ has been defined such that \eqref{another_r_c_condn} holds (see Section \ref{sec:parameters}).  Using the expansions \eqref{distance_expansion_eps} and \eqref{f_atm_expansion}, we find that the dimensionless model \eqref{dimensionless_model} becomes
\begin{subequations}
\label{dim_partial_exp_x}
\begin{align}
\rhovprimeprime
    & = ( \gv + \eps \g \rhov ) u^{ -3 / 2 } - f\atm( s ) b ( \rhovprime - \w )\\
	& = ( \gv + \eps \g \rhov ) \left[ 1 - 3 \eps ( \av \cdot \rhov ) + \O( \eps^2 ) \right] - \left[ f\atm( s_0 ) + \k \eps ( \av \cdot \rhov ) f\atm'( s_0 ) + \O( \eps^2 ) \right] b ( \rhovprime - \w )\\
	& =  \gv + \eps \g \rhov - 3 \gv \eps ( \av \cdot \rhov ) - f\atm( s_0 ) b ( \rhovprime - \w ) - \k \eps ( \av \cdot \rhov ) f\atm'( s_0 ) b ( \rhovprime - \w ) + \O( \eps^2 )\\
    & = \gv - f\atm( s_0 ) b ( \rhovprime - \w ) + \eps [ \g \rhov - 3 \gv ( \av \cdot \rhov ) - \k ( \av \cdot \rhov ) f\atm'( s_0 ) b ( \rhovprime - \w ) ] + \O( \eps^2 )\\
    & = \gv - b\eff ( \rhovprime - \w ) + \eps [ \g \rhov - 3 \gv ( \av \cdot \rhov ) - b\eff' ( \av \cdot \rhov ) ( \rhovprime - \w ) ] + \O( \eps^2 ),
\end{align}
\end{subequations}
where
\begin{subequations}
\begin{align}
b\eff  & \equiv f\atm( s_0 ) b,\\
b\eff' & \equiv \k f\atm'( s_0 ) b.
\end{align}
\end{subequations}
We think of $b\eff$ as the effective dimensionless linear drag coefficient at radius $R_0$, and we think of $b\eff'$ as the effective rate of change of $b\eff$ there.

We assume that a solution to the spatially expanded model \eqref{dim_partial_exp_x} with initial conditions \eqref{nondim_initial_conditions} exists and has the form
\bq
\label{ansatz}
\rhov = \rhov_0 + \eps \rhov_1 + \O( \eps^2 ).
\eq
Substituting \eqref{ansatz} into \eqref{dim_partial_exp_x}, we find that
\begin{empheq}[box=\mybox]{equation}
\label{nondim_ODE}
\rhovprimeprime_0 + \eps \rhovprimeprime_1 + \O( \eps^2 ) = \gv - b\eff ( \rhovprime_0 - \w ) + \eps [ \g \rhov_0 - 3 \gv ( \av \cdot \rhov_0 ) - b\eff' ( \av \cdot \rhov_0 ) ( \rhovprime_0 - \w ) - b\eff \rhovprime_1 ] + \O( \eps^2 ),
\end{empheq}
where we have collected the RHS in powers of $\eps$.

The initial conditions \eqref{nondim_initial_conditions} can similarly be written in powers of $\eps$, resulting in the initial conditions
\begin{subequations}
\label{ordered_initial_conditions}
\begin{empheq}[box=\mybox]{align}
\O( 1 ): \quad \rhov_0( 0 )	& = \vec{ 0 },\\
\rhovprime_0( 0 )				& = \dv,\\
\O( \eps ): \quad \rhov_1( 0 )	& = \vec{ 0 },\\
\rhovprime_1( 0 ) 				& = \vec{ 0 }.
\end{empheq}
\end{subequations}
The result \eqref{nondim_ODE} is an expansion of the dimensionless model \eqref{dimensionless_model} in terms of dimensionless functions $\rhov_j$, each corresponding to a power $j$ of the small, dimensionless parameter $\eps$.  We note that the coupled, nonlinear term $\F\grav$ appearing in the original model \eqref{original_model} is now approximated by de-coupled, linear functions in \eqref{nondim_ODE}, providing a significant simplification.  Similar comments apply to the atmospheric thinning function $f\atm$.

\section{Integral solutions for time-dependent wind}\label{sec:integral_solns}

We split our study of the IVP given by \eqref{nondim_ODE} and \eqref{ordered_initial_conditions} into two cases.\footnote{We split up the cases $b\eff = 0$ and $b\eff \ne 0$ because, for each $j$, in the $j\suptext{th}$-order problem, $b\eff$ multiplies $\rhovprime_j$; this causes the quantity $1 / b\eff$ to appear, which is singular when $b\eff = 0$.  There is an additional, third regime where $0 < b\eff \ll 1$ that becomes important in numerical implementations; see Section \ref{sec:b_eff_small} and Appendix \ref{sec:b_eff_expansions}.  We also note that the solutions for the case $b\eff = 0$ can be found from Taylor expansions of the solutions for the case $b\eff \ne 0$ (because the singularities mentioned above are removable), but we find it instructive to treat the two cases separately.}
\begin{enumerate}[label={(\textbf{\arabic*})}]

	\item The first case is defined by the condition
\bq
\label{bz}
b\eff = 0.\tag{\textbf{bz}}
\eq
The case \bz{} is that of quasi-negligible air resistance: $b\eff$ is negligible, due to a very small value of $b$ or to a very small value of $f\atm( s_0 )$ (or to a very small product of those factors), but $b\eff'$ has no such restriction, as the air resistance may be changing in a non-negligible way.

	\item The second case is defined by the condition
\bq
\label{bnz}
b\eff \ne 0\tag{\textbf{bnz}}
\eq
and accounts for cases of non-negligible air resistance.

\end{enumerate}

\subsection{Zeroth-order closed-form solution for $b\eff = 0$ and time-dependent wind}\label{sec:rho_0_bz}

To find the zeroth-order solution $\rhov_0( \tau ; \bz )$ of the Ansatz \eqref{ansatz}, we equate the $\O( 1 )$ terms of the ODE \eqref{nondim_ODE} after enforcing the condition \bz{} and include the $\O( 1 )$ initial conditions given by \eqref{ordered_initial_conditions} to find the IVP
\begin{subequations}
\label{rho_0_bz_IVP}
\begin{align}
\label{rho_0_bz_ODE}
\O( 1 ): \quad \rhovprimeprime_0( \tau ; \bz )	& = \gv,\\
\rhovprime_0( 0 ; \bz ) 							& = \dv,\\
\rhov_0( 0 ; \bz ) 									& = \vec{ 0 }.
\end{align}
\end{subequations}
Since we have enforced the condition \bz{} on the ODE \eqref{nondim_ODE} before seeking our solution, we've included that condition as a special argument of the solution, after the semicolon.  We then integrate \eqref{rho_0_bz_IVP} to write the zeroth-order solutions, for $0 \le \tau \le \tau_\star$, as
\begin{subequations}
\label{rho_0_bz}
\begin{empheq}[box=\mybox]{align}
\rhovprime_0( \tau ; \bz )	& = \dv + \gv \tau,\\
\rhov_0( \tau ; \bz ) 		& = \tau ( \dv + \tfrac{ 1 }{ 2 } \gv \tau ),
\end{empheq}
\end{subequations}
where $\tau_\star$, defined in \eqref{tau_star}, ensures the quality of our approximations.  Hence the scalar components of the vector-valued, zeroth-order solution $\rhov_0$ are completely de-coupled from one another.

\paragraph{Physical interpretation} The solution $\rhov_0$ is the dimensionless, scaled, translated, approximate solution to the physical problem modeled by \eqref{original_model}, neglecting the effects of air resistance (and therefore wind) and assuming the force of gravity to have constant strength $| \g |$ and fixed direction $-\av$ (pointing radially inward from the initial position of the projectile).

\subsection{First-order integral solution for $b\eff = 0$ and time-dependent wind}\label{sec:rho_1_bz}

Equating the $\O( \eps )$ terms of the ODE \eqref{nondim_ODE} and the $\O( \eps )$ initial conditions given by \eqref{ordered_initial_conditions}, we find the IVP
\begin{subequations}
\label{rho_1_bz_IVP}
\begin{align}
\label{rho_1_bz_ODE}
\O( \eps ): \quad \rhovprimeprime_1( \tau ; \bz )	& = \qv( \tau ; \bz ),\\
\rhovprime_1( 0 ; \bz ) 							& = \vec{ 0 },\\
\rhov_1( 0 ; \bz ) 									& = \vec{ 0 },
\end{align}
\end{subequations}
where
\bq
\qv( \tau ; \bz ) \equiv \g \rhov_0( \tau ; \bz ) - 3 [ \av \cdot \rhov_0( \tau ; \bz ) ] \gv - b\eff' [ \av \cdot \rhov_0( \tau ; \bz ) ] [ \rhovprime_0( \tau ; \bz ) - \w( \tau ) ].
\eq
Since the RHS of the ODE \eqref{rho_1_bz_ODE} contains only the known, zeroth-order solution $\rhov_0$, the scalar components of the unknown first-order solution $\rhov_1$ defined in \eqref{ansatz} have been effectively de-coupled.

Defining\footnote{We note the following physical significance of $\s$ when $v_0 \neq 0$: $\s = 1$ indicates radially outward initial motion along the $\av$-axis; $\s > 0$ indicates initial motion with both an outward and a tangential component; $\s = 0$ indicates initial motion perpendicular to the $\av$-axis (that is, initial motion tangential to the surface of the Earth); $\s < 0$ indicates initial motion with both an inward and a tangential component; $\s = -1$ indicates radially inward initial motion along the $\av$-axis.}
\bq
\s \equiv \av \cdot \bv
\eq
and using the fact that $| \av | = 1$, we find from \eqref{rho_0_bz} that
\bq
\label{alpha_dot_rho_0}
\av \cdot \rhov_0( \tau ; \bz ) = \tau ( \d \s + \tfrac{ 1 }{ 2 } \g \tau ).
\eq
We may then write the ODE \eqref{rho_1_bz_ODE} more explicitly as
\bq
\label{rhoprimeprime_1_bz}
\rhovprimeprime_1( \tau ; \bz ) = \vec{ c }_1 \tau + \vec{ c }_2 \tau^2 + \vec{ c }_3 \tau^3 + c_4( \tau ) \w( \tau ),
\eq
where
\begin{subequations}
\begin{align}
\vec{ c }_1 & \equiv \g \dv - 3 \d \s \gv - b\eff' \d \s \dv,\\
\vec{ c }_2 & \equiv -( \g \gv + b\eff' \d \s \gv + \tfrac{ 1 }{ 2 } b\eff' \g \dv ),\\
\vec{ c }_3 & \equiv -\tfrac{ 1 }{ 2 } b\eff' \g \gv,\\
c_4( \tau ) & \equiv b\eff' \tau ( \d \s + \tfrac{ 1 }{ 2 } \g \tau ).
\end{align}
\end{subequations}
Integrating \eqref{rhoprimeprime_1_bz} and using the initial conditions of \eqref{rho_1_bz_IVP}, we find, for $0 \le \tau \le \tau_\star$, that
\begin{subequations}
\label{rho_rhoprime_1_bz}
\begin{empheq}[box=\mybox]{equation}
\label{rhoprime_1_bz}
\rhovprime_1( \tau ; \bz ) = \tfrac{ 1 }{ 2 } \vec{ c }_1 \tau^2 + \tfrac{ 1 }{ 3 } \vec{ c }_2 \tau^3 + \tfrac{ 1 }{ 4 } \vec{ c }_3 \tau^4 + \int_0^\tau c_4( \tau' ) \w( \tau' ) d\tau'.
\end{empheq}
Integrating \eqref{rhoprime_1_bz}, we find that the solution to the $\O( \eps )$ IVP \eqref{rho_1_bz_IVP} can be written, for $0 \le \tau \le \tau_\star$, as
\begin{empheq}[box=\mybox]{equation}
\label{rho_1_bz}
\rhov_1( \tau ; \bz ) = \tfrac{ 1 }{ 6 } \vec{ c }_1 \tau^3 + \tfrac{ 1 }{ 12 } \vec{ c }_2 \tau^4 + \tfrac{ 1 }{ 20 } \vec{ c }_3 \tau^5 + \int_0^\tau \int_0^{ \tau' } c_4( \tau'' ) \w( \tau'' ) d\tau'' d\tau'.
\end{empheq}
\end{subequations}
\paragraph{Physical interpretation} To physically interpret the ODE \eqref{rho_1_bz_ODE}, we define
\bq
\rhov_0^{ \, \parallel } \equiv ( \av \cdot \rhov_0 ) \av
\eq
to be the component of $\rhov_0$ parallel to the $\av$-axis, and we define
\bq
\rhov_0^{ \, \perp } \equiv \rhov_0 - \rhov_0^{ \, \parallel }
\eq
to be the component of $\rhov_0$ perpendicular to the $\av$-axis.  We then write the ODE \eqref{rho_1_bz_ODE} as
\bq
\label{rho_1_bz_ODE_interp}
\rhovprimeprime_1 = 2 | \g | \rhov_0^{ \, \parallel } - | \g | \rhov_0^{ \, \perp } - b\eff' ( \av \cdot \rhov_0 ) ( \rhovprime_0 - \w ).
\eq
We interpret the first two terms of the RHS of \eqref{rho_1_bz_ODE_interp} to be a first correction to the $\O( 1 )$ dimensionless gravitational acceleration $\gv$ appearing on the RHS of \eqref{rho_0_bz_ODE}.  The first term in the correction, $2 | \g | \rhov_0^{ \, \parallel }$, is a radial correction; it nudges the projectile along the $\av$-axis a little farther from its initial position, in whichever direction along the $\av$-axis it was already moving (inward or outward); this is a first correction to the $\O( 1 )$ approximation that the gravitational force has constant \textsl{magnitude}.  The second term in the correction, $-| \g | \rhov_0^{ \, \perp }$, is a correction in the plane perpendicular to the $\av$-axis; it dampens the motion of the projectile in this plane, serving as a first correction to the $\O( 1 )$ approximation that the gravitational force has constant \textsl{direction}.\\

The final term on the RHS of \eqref{rho_1_bz_ODE_interp}, $-b\eff' ( \av \cdot \rhov_0 ) ( \rhovprime_0 - \w )$, is a first-order correction to the zeroth-order approximation of constant atmospheric density.  We note that this correction contains a factor of $\av \cdot \rhov_0$ because $f\atm$ is a function of the radial distance of the projectile from the center of the Earth.

\subsection{Two-term expansion for $b\eff = 0$ and time-dependent wind}\label{sec:solns_expanded_bz}

We define the two-term expansion $\rhov\comp$ to be the result of collecting and scaling the solutions given by \eqref{rho_0_bz} and \eqref{rho_1_bz}, with the expanded velocity defined similarly: for $0 \le \tau \le \tau_\star$,
\begin{subequations}
\begin{empheq}[box=\mybox]{align}
\rhov\comp( \tau ; \bz ) 		& \equiv \rhov_0( \tau ; \bz ) + \eps \rhov_1( \tau ; \bz ),\\
\rhovprime\comp( \tau ; \bz )	& \equiv \rhovprime_0( \tau ; \bz ) + \eps \rhovprime_1( \tau ; \bz ).
\end{empheq}
\end{subequations}
The two-term expansion $\rhov\comp$ is the dimensionless, scaled, translated, approximate solution to the problem modeled by \eqref{dim_partial_exp_x} in the case of quasi-negligible air resistance \bz, with the force of gravity spatially linearized about the initial position $\absposv_0$ of \eqref{original_initial_conditions}.

\subsection{Zeroth-order integral solution for $b\eff \ne 0$ and time-dependent wind}\label{sec:rho_0_bnz}

To find the zeroth-order solution $\rhov_0( \tau ; \bnz )$ of the Ansatz \eqref{ansatz}, we equate the $\O( 1 )$ terms of the ODE \eqref{nondim_ODE} after enforcing the condition \bnz{} and include the $\O( 1 )$ initial conditions given by \eqref{ordered_initial_conditions} to find the IVP
\begin{subequations}
\label{rho_0_bnz_IVP}
\begin{align}
\label{rho_0_bnz_ODE}
\O( 1 ): \quad \rhovprimeprime_0( \tau ; \bnz )	& = \gv - b\eff [ \rhovprime_0( \tau ; \bnz ) - \w( \tau ) ],\\
\rhovprime_0( 0 ; \bnz ) 							& = \dv,\\
\rhov_0( 0 ; \bnz ) 								& = \vec{ 0 }.
\end{align}
\end{subequations}
Define
\bq
\label{tau_decay_defn}
\td \equiv \frac{ 1 }{ b\eff }
\eq
to be the time scale over which the initial velocity of the projectile decays.  Then the solution to the IVP \eqref{rho_0_bnz_IVP} is
\begin{subequations}
\label{solns_0_bnz}
\begin{empheq}[box=\mybox]{equation}
\label{rho_0_bnz}
\rhov_0( \tau ; \bnz ) \equiv \td \left[ \gv \tau + ( \dv - \td \gv ) ( 1 - \expmt ) \right] + b\eff I_1[ \w ]( \tau ),
\end{empheq}
with velocity
\begin{empheq}[box=\mybox]{equation}
\label{rhoprime_0_bnz}
\rhovprime_0( \tau ; \bnz ) \equiv \dv \expmt + \td \gv ( 1 - e^{ -\tau / \td } ) + b\eff \expmt \I{ 0 }{ \w }( \tau ),
\end{empheq}
\end{subequations}
where, for $n \in \{ 0 , 1 \}$ and for $h$ a function of a single variable, we define
\bq
\label{I_defn}
\I{ n }{ h }( \tau ) \equiv \bit{ n = 0 } \int_0^\tau e^{ \tau' / \td } h( \tau' ) \, d\tau' + \bit{ n = 1 } \int_0^\tau e^{ -\tau' / \td } \I{ 0 }{ h }( \tau' ) \, d\tau'.
\eq
In the event that $h$ is vector-valued, we apply $I_n$ component-wise.

\paragraph{Physical interpretation} The ODE \eqref{rho_0_bnz_ODE} is a direct approximation of the original model \eqref{original_model} in dimensionless form, with the simplifying assumptions of a constant gravitational force and constant atmospheric thinning factor.

\subsection{First-order integral solution for $b\eff \ne 0$ and time-dependent wind}\label{sec:rho_1_bnz}

Equating the $\O( \eps )$ terms of the ODE \eqref{nondim_ODE} and the $\O( \eps )$ initial conditions given by \eqref{ordered_initial_conditions}, we find the IVP
\begin{subequations}
\begin{align}
\label{rho_1_bnz_ODE}
\O( \eps ): \quad \rhovprimeprime_1( \tau ; \bnz )	& = \qv( \tau ; \bnz ) - b\eff \rhovprime_1( \tau ; \bnz ),\\
\rhovprime_1( 0 ; \bnz ) 								& = \vec{ 0 },\\
\rhov_1( 0 ; \bnz ) 									& = \vec{ 0 },
\end{align}
\end{subequations}
where
\bq
\qv( \tau ; \bnz ) \equiv \g \rhov_0( \tau ; \bnz ) - 3 [ \av \cdot \rhov_0( \tau ; \bnz ) ] \gv - b\eff' [ \av \cdot \rhov_0( \tau ; \bnz ) ] [ \rhovprime_0( \tau ; \bnz ) - \w( \tau ) ].
\eq
The solution to this IVP is
\begin{subequations}
\label{solns_1_bnz}
\begin{empheq}[box=\mybox]{equation}
\label{rho_1_bnz}
\rhov_1( \tau ; \bnz ) = \I{ 1 }{ x \mapsto \qv( x ; \bnz ) \, }( \tau ),
\end{empheq}
with velocity
\begin{empheq}[box=\mybox]{equation}
\rhovprime_1( \tau ; \bnz ) = \expmt \I{ 0 }{ x \mapsto \qv( x ; \bnz ) \, }( \tau ).
\end{empheq}
\end{subequations}
We calculate
\begin{align}
\I{ n }{ x \mapsto \qv( x ; \bnz ) }( \tau )
	& = \g \I{ n }{ x \mapsto \rhov_0( x ; \bnz ) }( \tau ) - 3 [ \av \cdot \I{ n }{ x \mapsto \rhov_0( x ; \bnz ) }( \tau ) ] \gv\\
	\nonumber%
	& \quad - b\eff' \I{ n }{ x \mapsto [ \av \cdot \rhov_0( x ; \bnz ) ] \rhovprime_0( x ; \bnz ) }( \tau )\\
	\nonumber%
	& \quad + b\eff' \I{ n }{ x \mapsto [ \av \cdot \rhov_0( x ; \bnz ) ] \w( x ) }( \tau ),
\end{align}
\begin{align}
\I{ n }{ x \mapsto \rhov_0( x ; \bnz ) }( \tau )
	& = \td ( \dv - \td \gv ) \I{ n }{ x \mapsto 1 }( \tau ) + \td \gv \I{ n }{ x \mapsto x }( \tau )\\
	\nonumber%
	& \quad - \td ( \dv - \td \gv ) \I{ n }{ x \mapsto e^{ -x / \td } }( \tau ) + b\eff \I{ n }{ x \mapsto \I{ 1 }{ \w }( x ) }( \tau ),
\end{align}
and
\begin{align}
& \I{ n }{ x \mapsto [ \av \cdot \rhov_0( x ; \bnz ) ] \rhovprime_0( x ; \bnz ) }( \tau )\\
	\nonumber%
	& = \I{ n }{ x \mapsto \vec{ \Omega }( x ; \bnz ) }( \tau ) + \td^2 ( \d \s - \td \g ) \gv \I{ n }{ x \mapsto 1 }( \tau ) + \td^2 \g \gv \I{ n }{ x \mapsto x }( \tau )\\
	\nonumber%
	& \quad + \td ( \d \s - \td \g ) ( \dv - 2 \td \gv ) \I{ n }{ x \mapsto e^{ -x / \td } }( \tau ) + \td \g ( \dv - \td \gv ) \I{ n }{ x \mapsto x e^{ -x / \td } }( \tau )\\
	\nonumber%
	& \quad - \td ( \d \s - \td \g ) ( \dv - \td \gv ) \I{ n }{ x \mapsto e^{ -2 x / \td } }( \tau ),
\end{align}
where
\begin{align}
\vec{ \Omega }( \tau ; \bnz )
	& \equiv b\eff \I{ 1 }{ \av \cdot \w }( \tau ) [ \td \gv + ( \dv - \td \gv ) \expmt + b\eff \expmt \I{ 0 }{ \w }( \tau ) ]\\
	\nonumber%
	& \quad + \expmt \I{ 0 }{ \w }( \tau ) [ \g \tau + ( \d \s - \td \g ) ( 1 - \expmt ) + b\eff^2 \I{ 1 }{ \av \cdot \w }( \tau ) ],
\end{align}
with selected integrals $I_n$ provided in Appendix \ref{sec:I_n}.  The results above and in Appendix \ref{sec:I_n}, combined with linearity, provide integral representations of the solutions \eqref{solns_1_bnz} in terms of the wind function $\w$.

\paragraph{Physical interpretation} To physically interpret the ODE \eqref{rho_1_bnz_ODE}, we first note that $\qv( \tau ; \bnz )$ can be understood in the same way as $\qv( \tau ; \bz )$.  The remaining term, $-b\eff \rhovprime_1( \tau ; \bnz )$, is the $\O( \eps )$ force of linear air resistance, taking into account the Ansatz \eqref{ansatz}; it contains no wind term because the wind is assumed to be of $\O( 1 )$.

\subsection{Two-term expansion for $b\eff \ne 0$ and time-dependent wind}\label{sec:solns_expanded_bnz}

We define the two-term expansion $\rhov\comp$ to be the result of collecting and scaling the solutions given by \eqref{rho_0_bnz} and \eqref{rho_1_bnz}, with the expanded velocity $\rhovprime\comp$ defined similarly: for $0 \le \tau \le \tau_\star$,
\begin{subequations}
\begin{empheq}[box=\mybox]{align}
\rhov\comp( \tau ; \bnz )		& \equiv \rhov_0( \tau ; \bnz ) + \eps \rhov_1( \tau ; \bnz ),\\
\rhovprime\comp( \tau ; \bnz )	& \equiv \rhovprime_0( \tau ; \bnz ) + \eps \rhovprime_1( \tau ; \bnz ).
\end{empheq}
\end{subequations}
The two-term expansion $\rhov\comp$ is the dimensionless, scaled, translated, approximate solution to the problem modeled by \eqref{dim_partial_exp_x}, in the case of non-negligible air resistance \bnz, with the force of gravity spatially linearized about the initial position $\absposv_0$ of \eqref{original_initial_conditions}.

\section{Closed-form solutions for constant wind}\label{sec:const_wind}

We write the special argument for this section as \cw, given by the condition
\bq
\label{cw}
\w( \tau ) \equiv \w_0 ~ \exists ~ \text{fixed} ~ \w_0 \in \mathbb{ R }^3\tag{\textbf{cw}}.
\eq

\subsection{Zeroth-order closed-form solution for $b\eff = 0$ and constant wind}\label{sec:rho_0_bz_cw}

The solutions \eqref{rho_0_bz} are independent of wind and therefore unaffected by the condition \cw; we find that
\begin{subequations}
\label{solns_0_bz_cw}
\begin{empheq}[box=\mybox]{align}
\rhov_0( \tau ; \bz , \cw ) 		& = \rhov_0( \tau ; \bz ),\\
\rhovprime_0( \tau ; \bz , \cw )	& = \rhovprime_0( \tau ; \bz ),
\end{empheq}
\end{subequations}
both valid for $0 \le \tau \le \tau_\star$.

\subsection{First-order closed-form solution for $b\eff = 0$ and constant wind}

Evaluating the solutions \eqref{rho_rhoprime_1_bz} using the condition \cw, we find that% \comment{see General atm fcn 2023-07-01}
\begin{subequations}
\label{solns_1_bz_cw}
\begin{empheq}[box=\mybox]{align}
\rhovprime_1( \tau ; \bz , \cw )	& = \tfrac{ 1 }{ 2 } \vec{ c }_1 \tau^2 + \tfrac{ 1 }{ 3 } \vec{ c }_2 \tau^3 + \tfrac{ 1 }{ 4 } \vec{ c }_3 \tau^4 + b\eff' \w_0 ( \tfrac{ 1 }{ 2 } \d \s \tau^2 + \tfrac{ 1 }{ 6 } \g \tau^3 ),\\
\rhov_1( \tau ; \bz , \cw ) 		& = \tfrac{ 1 }{ 6 } \vec{ c }_1 \tau^3 + \tfrac{ 1 }{ 12 } \vec{ c }_2 \tau^4 + \tfrac{ 1 }{ 20 } \vec{ c }_3 \tau^5 + b\eff' \w_0 ( \tfrac{ 1 }{ 6 } \d \s \tau^3 + \tfrac{ 1 }{ 24 } \g \tau^4 ),
\end{empheq}
\end{subequations}
both valid for $0 \le \tau \le \tau_\star$.

\subsection{Two-term expansion for $b\eff = 0$ and constant wind}\label{sec:solns_expanded_bz_cw}

We define the two-term expansion $\rhov\comp$ to be the result of collecting and scaling the solutions given by \eqref{solns_0_bz_cw} and \eqref{solns_1_bz_cw}, with the expanded velocity defined similarly: for $0 \le \tau \le \tau_\star$,
\begin{subequations}
\begin{empheq}[box=\mybox]{align}
\rhov\comp( \tau ; \bz , \cw )			& \equiv \rhov_0( \tau ; \bz , \cw ) + \eps \rhov_1( \tau ; \bz , \cw ),\\
\rhovprime\comp( \tau ; \bz , \cw )	& \equiv \rhovprime_0( \tau ; \bz , \cw ) + \eps \rhovprime_1( \tau ; \bz , \cw ).
\end{empheq}
\end{subequations}
The two-term expansion $\rhov\comp$ is the dimensionless, scaled, translated, approximate solution to the problem modeled by \eqref{dim_partial_exp_x} in the case of quasi-negligible air resistance and constant wind, with the force of gravity spatially linearized about the initial position $\absposv_0$ of \eqref{original_initial_conditions}.

\subsection{Zeroth-order closed-form solution for $b\eff \ne 0$ and constant wind}\label{sec:rho_0_bnz_cw}

Evaluating the solutions \eqref{solns_0_bnz} using the condition \cw, we find that
\begin{subequations}
\begin{empheq}[box=\mybox]{align}
\label{rho_0_bnz_cw}
\rhov_0( \tau ; \bnz , \cw ) & = \Cv \tau + \td \Dv ( 1 - \expmt ),\\
\label{rhoprime_0_bnz_cw}
\rhovprime_0( \tau ; \bnz , \cw ) & = \Cv + \expmt \Dv,
\end{empheq}
\end{subequations}
where
\begin{subequations}
\begin{align}
\Cv & \equiv \w_0 + \td \gv,\\
\Dv & \equiv \dv - \Cv.
\end{align}
\end{subequations}

\subsection{First-order closed-form solution for $b\eff \ne 0$ and constant wind}

Evaluating the solutions \eqref{solns_1_bnz} using the condition \cw, we find that
\begin{subequations}
\label{solns_1_bnz_cw}
\begin{empheq}[box=\mybox]{align}
\label{rhov_1_bnz_cw}
\rhov_1( \tau ; \bnz , \cw ) 		& = \I{ 1 }{ x \mapsto \qv( x ; \bnz , \cw ) \, }( \tau ),\\
\label{rhovprime_1_bnz_cw}
\rhovprime_1( \tau ; \bnz , \cw )	& = \expmt \I{ 0 }{ x \mapsto \qv( x ; \bnz , \cw ) \, }( \tau ),
\end{empheq}
\end{subequations}
where
\begin{align}
\qv( \tau ; \bnz , \cw ) 	& \equiv \g \rhov_0( \tau ; \bnz , \cw ) - 3 [ \av \cdot \rhov_0( \tau ; \bnz , \cw ) ] \gv\\
							\nonumber%
							& \quad - b\eff' [ \av \cdot \rhov_0( \tau ; \bnz , \cw ) ] [ \rhovprime_0( \tau ; \bnz , \cw ) - \w_0 ].
\end{align}
We calculate
\begin{align}
\label{I_n_q_cw}
\I{ n }{ x \mapsto \qv( x ; \bnz , \cw ) }( \tau )
	& = \g \I{ n }{ x \mapsto \rhov_0( x ; \bnz , \cw ) }( \tau )\\
	\nonumber%
	& \quad - 3 [ \av \cdot \I{ n }{ x \mapsto \rhov_0( x ; \bnz , \cw ) }( \tau ) ] \gv\\
	\nonumber%
	& \quad - b\eff' \I{ n }{ x \mapsto [ \av \cdot \rhov_0( x ; \bnz , \cw ) ] \rhovprime_0( x ; \bnz , \cw ) }( \tau )\\
	\nonumber%
	& \quad + b\eff' \I{ n }{ x \mapsto [ \av \cdot \rhov_0( x ; \bnz , \cw ) ] \w_0 }( \tau ),
\end{align}
\begin{subequations}
\begin{align}
\I{ n }{ x \mapsto [ \av \cdot \rhov_0( x ; \bnz , \cw ) ] \w_0 }( \tau )
	& = \I{ n }{ x \mapsto \td ( \av \cdot \Dv ) \w_0 + ( \av \cdot \Cv ) \w_0 x - \td ( \av \cdot \Dv ) \w_0 e^{ -x / \td } }( \tau )\\
	& = \td ( \av \cdot \Dv ) \w_0 \I{ n }{ x \mapsto 1 }( \tau ) + ( \av \cdot \Cv ) \w_0 \I{ n }{ x \mapsto x }( \tau )\\
	\nonumber%
	& \quad - \td ( \av \cdot \Dv ) \w_0 \I{ n }{ x \mapsto e^{ -x / \td } }( \tau ),
\end{align}
\end{subequations}
\begin{align}
\I{ n }{ x \mapsto \rhov_0( x ; \bnz , \cw ) }( \tau )
	& = \td ( \dv - \td \gv ) \I{ n }{ x \mapsto 1 }( \tau ) + \td \gv \I{ n }{ x \mapsto x }( \tau )\\
	\nonumber%
	& \quad - \td ( \dv - \td \gv ) \I{ n }{ x \mapsto e^{ -x / \td } }( \tau ) + b\eff \I{ n }{ x \mapsto \I{ 1 }{ y \mapsto \w_0 }( x ) }( \tau ),
\end{align}
\begin{subequations}
\begin{align}
\I{ n }{ x \mapsto \I{ 1 }{ y \mapsto \w_0 }( x ) }( \tau )
	& = \w_0 \I{ n }{ x \mapsto \I{ 1 }{ y \mapsto 1 }( x ) }( \tau )\\
	& = \w_0 [ -\td^2 \I{ n }{ x \mapsto 1 }( \tau ) + \td \I{ n }{ x \mapsto x }( \tau ) + \td^2 \I{ n }{ x \mapsto e^{ -x / \td } }( \tau ) ],
\end{align}
\end{subequations}
and
\begin{align}
& \I{ n }{ x \mapsto [ \av \cdot \rhov_0( x ; \bnz , \cw ) ] \rhovprime_0( x ; \bnz , \cw ) }( \tau )\\
	\nonumber%
	& = \I{ n }{ x \mapsto \vec{ \Omega }( x ; \bnz , \cw ) }( \tau ) + \td^2 ( \d \s - \td \g ) \gv \I{ n }{ x \mapsto 1 }( \tau ) + \td^2 \g \gv \I{ n }{ x \mapsto x }( \tau )\\
	\nonumber%
	& \quad + \td ( \d \s - \td \g ) ( \dv - 2 \td \gv ) \I{ n }{ x \mapsto e^{ -x / \td } }( \tau ) + \td \g ( \dv - \td \gv ) \I{ n }{ x \mapsto x e^{ -x / \td } }( \tau )\\
	\nonumber%
	& \quad - \td ( \d \s - \td \g ) ( \dv - \td \gv ) \I{ n }{ x \mapsto e^{ -2 x / \td } }( \tau ),
\end{align}
where
\begin{align}
\label{W_lw}
\vec{ \Omega }( \tau ; \bnz , \cw )
	& \equiv \td [ ( \av \cdot \Dv ) \w_0 - ( \av \cdot \w_0 ) \Cv ] + [ ( \av \cdot \w_0 ) \Cv + ( \av \cdot \Cv ) \w_0 ] \tau\\
	\nonumber%
	& \quad + \td [ ( \av \cdot \w_0 ) \Cv - ( \av \cdot \w_0 ) \Dv - 2 \w_0 ( \av \cdot \Dv ) ] \expmt + [ ( \av \cdot \w_0 ) \Dv - ( \av \cdot \Cv ) \w_0 ] \tau \expmt\\
	\nonumber%
	& \quad + \td [ ( \av \cdot \w_0 ) \Dv + ( \av \cdot \Dv ) \w_0 ] e^{ -2 \tau / \td }
\end{align}
and
\begin{align}
\label{I_n_W}
\I{ n }{ x \mapsto \vec{ \Omega }( x ; \bnz , \cw ) }( \tau )
	& = \td [ ( \av \cdot \Dv ) \w_0 - ( \av \cdot \w_0 ) \Cv ] \I{ n }{ x \mapsto 1 }( \tau )\\
	\nonumber%
	& \quad + [ ( \av \cdot \w_0 ) \Cv + ( \av \cdot \Cv ) \w_0 ] \I{ n }{ x \mapsto x }( \tau )\\
	\nonumber%
	& \quad + \td [ ( \av \cdot \w_0 ) \Cv - ( \av \cdot \w_0 ) \Dv - 2 \w_0 ( \av \cdot \Dv ) ] \I{ n }{ x \mapsto e^{ -x / \td } }( \tau )\\
	\nonumber%
	& \quad + [ ( \av \cdot \w_0 ) \Dv - ( \av \cdot \Cv ) \w_0 ] \I{ n }{ x \mapsto x e^{ -x / \td } }( \tau )\\
	\nonumber%
	& \quad + \td [ ( \av \cdot \w_0 ) \Dv + ( \av \cdot \Dv ) \w_0 ] \I{ n }{ x \mapsto e^{ -2 x / \td } }( \tau ). 
\end{align}
The results of this section, together with linearity and the calculations of the integrals
\[
\I{ n }{ x \mapsto \{ 1 , x , e^{ -x / \td } , x e^{ -x / \td } , e^{ -2 x / \td } \} }( \tau )
\]
found in Appendix \ref{sec:I_n}, provide a closed-form representation of the solutions \eqref{solns_1_bnz_cw}.

\subsection{Two-term expansion for $b\eff \ne 0$ and constant wind}\label{sec:solns_expanded_bnz_cw}

We define the two-term expansion $\rhov\comp$ to be the result of collecting and scaling the solutions given by \eqref{rho_0_bnz_cw} and \eqref{rhov_1_bnz_cw}, with the expanded velocity $\rhovprime\comp$ defined similarly: for $0 \le \tau \le \tau_\star$,
\begin{subequations}
\begin{empheq}[box=\mybox]{align}
\rhov\comp( \tau ; \bnz , \cw ) 		& \equiv \rhov_0( \tau ; \bnz , \cw ) + \eps \rhov_1( \tau ; \bnz , \cw ),\\
\rhovprime\comp( \tau ; \bnz , \cw ) 	& \equiv \rhovprime_0( \tau ; \bnz , \cw ) + \eps \rhovprime_1( \tau ; \bnz , \cw ).
\end{empheq}
\end{subequations}
The two-term expansion $\rhov\comp$ is the dimensionless, scaled, translated, approximate solution to the problem modeled by \eqref{dim_partial_exp_x}, in the case of non-negligible air resistance \bnz{} and constant wind \cw, with the force of gravity spatially linearized about the initial position $\absposv_0$ of \eqref{original_initial_conditions}.

\section{Solutions of the original projectile-motion model}\label{sec:solns_to_original}

In Sections \ref{sec:integral_solns} and \ref{sec:const_wind} we provided various solutions to the dimensionless version of our problem, as introduced in Sections \ref{sec:model_tfn} and \ref{sec:spatial_expn}.  We now discuss the transformations of these dimensionless solutions back to the original time and distance scales, producing solutions of the original model \eqref{original_model} with initial conditions \eqref{original_initial_conditions}.

\subsection{Zeroth-order solutions of the original projectile-motion model}

Using the definition \eqref{relpos_defn} and scaling equations \eqref{scaling_eqn}, we find that each zeroth-order position (and its zeroth-order velocity) -- and approximate solution of the original, physically motivated system \eqref{original_model} with initial conditions \eqref{original_initial_conditions} -- can be written, for $0 \le t \le t\c \tau_\star$, in the form
\begin{subequations}
\begin{empheq}[box=\mybox]{align}
\absposv\subtext{order 0}( t ; ( \ldots ) ) 		& \equiv \absposv_0 + r\c \rhov_0\left( \tau = \frac{ t }{ t\c } ; ( \ldots ) \right),\\
\absposvprime\subtext{order 0}( t ; ( \ldots ) )	& \equiv v\c \rhovprime_0\left( \tau = \frac{ t }{ t\c } ; ( \ldots ) \right),
\end{empheq}
\end{subequations}
where $( \ldots )$ represents any special arguments inherited from the solutions $\rhov_0$ and $\rhovprime_0$.  In this way, we can find $\absposv\subtext{order 0}( t ; \bz )$ and $\absposvprime\subtext{order 0}( t ; \bz )$ from the solutions of Section \ref{sec:rho_0_bz}, $\absposv\subtext{order 0}( t ; \bnz )$ and $\absposvprime\subtext{order 0}( t ; \bnz )$ from the solutions of Section \ref{sec:rho_0_bnz}, $\absposv\subtext{order 0}( t ; \bz , \cw )$ and $\absposvprime\subtext{order 0}( t ; \bz , \cw )$ from the solutions of Section \ref{sec:rho_0_bz_cw}, and $\absposv\subtext{order 0}( t ; \bnz , \cw )$ and $\absposvprime\subtext{order 0}( t ; \bnz , \cw )$ from the solutions of Section \ref{sec:rho_0_bnz_cw}.

Each zeroth-order solution $\absposv\subtext{order 0}$ is the exact solution to the approximation of \eqref{original_model} found via spatial linearization, up to $\O( 1 )$, under the condition(s) indicated in its special argument(s).  In Section \ref{sec:model_validity}, we discuss the limitations of our spatial linearization approximations.

\subsection{Expanded solutions of the original projectile-motion model}

In the same way, we find that each expanded position (and its expanded velocity) -- and approximate solution of the original, physically motivated system \eqref{original_model} with initial conditions \eqref{original_initial_conditions} -- can be written, for $0 \le t \le t\c \tau_\star$, in the form
\begin{subequations}
\begin{empheq}[box=\mybox]{align}
\absposv\comp( t ; ( \ldots ) ) 		& \equiv \absposv_0 + r\c \rhov\comp\left( \tau = \frac{ t }{ t\c } ; ( \ldots ) \right),\\
\absposvprime\comp( t ; ( \ldots ) )	& \equiv v\c \rhovprime\comp\left( \tau = \frac{ t }{ t\c } ; ( \ldots ) \right),
\end{empheq}
\end{subequations}
where $( \ldots )$ represents any special arguments inherited from the solutions $\rhov\comp$ and $\rhovprime\comp$.  In this way, we can find $\absposv\comp( t ; \bz )$ and $\absposvprime\comp( t ; \bz )$ from the solutions of Section \ref{sec:solns_expanded_bz}, $\absposv\comp( t ; \bnz )$ and $\absposvprime\comp( t ; \bnz )$ from the solutions of Section \ref{sec:solns_expanded_bnz}, $\absposv\comp( t ; \bz , \cw )$ and $\absposvprime\comp( t ; \bz , \cw )$ from the solutions of Section \ref{sec:solns_expanded_bz_cw}, and $\absposv\comp( t ; \bnz , \cw )$ and $\absposvprime\comp( t ; \bnz , \cw )$ from the solutions of Section \ref{sec:solns_expanded_bnz_cw}.

Each expanded solution $\absposv\comp$ is the exact solution to the approximation of \eqref{original_model} found via spatial linearization, up to $\O( \eps )$, under the condition(s) indicated in its special argument(s).  In Section \ref{sec:model_validity}, we discuss the limitations of our spatial linearization approximations.

\section{Model parameters and validity}\label{sec:model_parameters_and_validity}

Here we define the model parameters $r\c$ and $t\c$ introduced in Section \ref{sec:model_tfn}.  We then discuss the temporal region of validity of the perturbation solutions provided in Sections \ref{sec:integral_solns} and \ref{sec:const_wind} and in Appendix \ref{sec:b_eff_expansions}.

\subsection{Choosing the parameters $r\c$ and $t\c$}\label{sec:parameters}

In Section \ref{sec:model_tfn}, we assumed that we can define values $0 < t\c$ and $0 < r\c$ such that $r\c \ll R_0$ (equivalent to $\eps \ll 1$); the quality of the resulting solutions depends on the assumptions $\eta \ll 1$ and $\eps \ll 1$.  In Section \ref{sec:spatial_expn}, we also assumed that we could define $r\c$ such that $\k \eps \rho\submax \ll 1$; the validity of the spatial expansion of $f\atm$ depends on this assumption.  For this reason, we need a method to guarantee these conditions.

\subsubsection{Bounding $r\c$ based on $\eta$}

To justify truncating \eqref{distance_expansion}, we require that $\eta \ll 1$.  To accomplish this, we develop conditions on $r\c$ that guarantee $\eta \le \eta\submax$ for some error-control parameter $\eta\submax$ satisfying
\bq
\label{eta_max_condition}
0 < \eta\submax \ll 1.
\eq
From \eqref{eta_u_defn} and the fact that $\av$ is a unit vector, we find that
\begin{subequations}
\begin{align}
\eta
 & = 2 \eps ( \av \cdot \rhov ) + \eps^2 \rho^2\\
 \label{eta_ineq}
 & \le 2 \eps \rho + \eps^2 \rho^2.
\end{align}
\end{subequations}
Requiring $\eqref{eta_ineq} \le \eta\submax$ then implies that
\bq
\label{eps_condition}
\eps \rho \le \sqrt{ 1 + \eta\submax } - 1.
\eq
Assuming that there is some parameter $\rho\submax$ such that\footnote{See Section \ref{sec:model_validity} for a discussion of the parameter $\rho\submax$.}
\bq
\label{rho_max_condition}
0 < \rho\submax \quad \text{and} \quad \rho( \tau ) \le \rho\submax ~ \text{for each value of $\tau$ at which $\rho$ is evaluated},
\eq
then the condition \eqref{eps_condition} is guaranteed if
\bq
\label{eps_condition_rho}
\eps \le \frac{ 1 }{ \rho\submax } \left[ \sqrt{ 1 + \eta\submax } - 1 \right];
\eq
from the definition \eqref{eps_defn}, we then find the requirement that
\bq
\label{r_c_condition_eta}
r\c \le \frac{ R_0 }{ \rho\submax } \left[ \sqrt{ 1 + \eta\submax } - 1 \right].
\eq
As long as $\eta\submax$ satisfies \eqref{eta_max_condition} and $\rho\submax$ satisfies \eqref{rho_max_condition}, then any value of $r\c$ satisfying \eqref{r_c_condition_eta} guarantees that $\eta \ll 1$.  In this case, our expansion \eqref{distance_expansion_eta} in terms of $\eta$ is justified.

\subsubsection{Bounding $r\c$ based on $\eps$}

To justify our solutions in powers of $\eps$, we require that $\eps \ll 1$, or, equivalently, that $r\c / R_0 \ll 1$.  Thus if we want to bound $\eps$ by requiring that $\eps \le \eps\submax ~ \exists ~ 0 < \eps\submax \ll 1$ chosen as an error-control parameter, then this is guaranteed if $r\c / R_0 \le \eps\submax$, or
\bq
\label{r_c_condition_eps}
r\c \le \eps\submax R_0.
\eq

\subsubsection{Bounding $r\c$ based on the spatial expansion of $f\atm$}

Looking again at the condition \eqref{another_r_c_condn},
\bq
\frac{ r\c \rho\submax }{ \ell } \ll 1,
\eq
for the spatial expansion of $f\atm$ to be valid, we then require that $\text{LHS} \le \nu\submax ~ \exists ~ 0 < \nu\submax \ll 1$ chosen as an error-control parameter, or
\bq
\label{r_c_condition_f_atm}
r\c \le \frac{ \ell \nu\submax }{ \rho\submax }.
\eq

\subsubsection{Defining $r\c$}

Putting the conditions \eqref{r_c_condition_eta}, \eqref{r_c_condition_eps}, and \eqref{r_c_condition_f_atm} together, we define
\begin{empheq}[box=\mybox]{equation}
\label{r_c_defn}
r\c \equiv \min\left\{ \frac{ R_0 }{ \rho\submax } \left[ \sqrt{ 1 + \eta\submax } - 1 \right] , \eps\submax R_0 , \frac{ \ell \nu\submax }{ \rho\submax } \right\},
\end{empheq}
where we require only that $0 < \rho\submax$, $0 < \eta\submax \ll 1$, $0 < \eps\submax \ll 1$, and $0 < \nu\submax \ll 1$.  In Section \ref{sec:traj_plots} we discuss the choices of the values of these error-control parameters.

\subsubsection{Defining $t\c$}

It would be natural to define the characteristic time $t\c$ to be $r\c / v_0$, consistent with \cite[p.~2]{Holmes}, but we'd like to permit values of the initial speed $v_0$ that are small, or even zero; instead, we define% \comment{what about $t\c = \td$ when $b\eff \neq 0$?  it \textbf{is} a characteristic, dimensionless time scale...}
\begin{empheq}[box=\mybox]{equation}
\label{t_c_defn}
t\c \equiv \frac{ r\c }{ \max\{ 1 , v_0 \} }.
\end{empheq}
This ensures that $t\c$ is well-defined -- both mathematically and numerically -- for $0 \le v_0$, while still providing the intended scaling for many values of $v_0$.

\subsection{Temporal region of validity of the model and its solutions}\label{sec:model_validity}

The condition \eqref{rho_max_condition} requires the value of $\rho( \tau )$ to never exceed the value of the error-control parameter $\rho\submax$.  When computing a quantity at dimensionless time $\tau$, then, one must verify that
\bq
\label{rho_valid_condition}
\rho( \tau' ) \le \rho\submax ~ \forall ~ \tau' \in [ 0 , \tau ].
\eq
This is because our solutions arise via integrals of the form $\int_0^\tau f( \tau' ) \, d\tau'$, which carry the implicit assumption that the integrand is valid at all values of $\tau' \in [ 0 , \tau ]$; regions of this closed interval for which $\rho( \tau' ) > \rho\submax$ indicate regions where the condition \eqref{rho_max_condition} is violated.  In these regions, the condition \eqref{r_c_condition_eta} is suspect, and so the requirement $\eta \ll 1$ is also suspect, putting our expansion in terms of $\eta$ in jeopardy; in order to guarantee that $\eta \ll 1$ for a particular value of $\tau$, then, one must ensure that \eqref{rho_valid_condition} holds.

We define the quantity
\bq
\label{tau_star}
\tau_\star \equiv \max\{ \tau \mid \eqref{rho_valid_condition} \text{ holds} \};
\eq
with this definition, the model and subsequent solutions developed in this document are valid for any value of $\tau$ satisfying
\bq
\label{tau_valid_condition}
0 \le \tau \le \tau_\star.
\eq
Below are three ways to ensure that the values of $\tau$ used for computational purposes satisfy \eqref{tau_valid_condition}.
\begin{enumerate}[label={(\textbf{\arabic*})}]

	\item One can begin at $\tau = 0$ and increment by some $\Delta \tau$ until the condition \eqref{rho_valid_condition} is violated.
	
	\item One can use a root-finding procedure to determine $\tau_\star$ from the condition \eqref{rho_valid_condition}.
	
	\item One can bound $\tau_\star$ from below by some quantity $\tau\submax\safe$, creating an interval $[ 0 , \tau\submax\safe ]$ within which the solutions can always be safely evaluated.

\end{enumerate}
We focus on strategy (\textbf{3}) and specialize our development to the constant-wind case \cw.  To find a value $\tau\submax\safe$ guaranteed to satisfy
\bq
\label{tau_desired_condition}
0 \le \tau\submax\safe \le \tau_\star,
\eq
we consider the speed of the projectile.  The projectile's speed can be increased only by the force of gravity or by the force of wind (if the object is at rest or has a tailwind).  The magnitude of the acceleration of gravity is bounded above by $g$, and the magnitude of the acceleration due to a tailwind is bounded above by $\tfrac{ 1 }{ m } b\drag | \vwind |$.  Defining
\bq
\hat{ g } \equiv g + \tfrac{ 1 }{ m } b\drag | \vwind |,
\eq
we find that the projectile's speed $v = | d\absposv / dt |$ satisfies, for $0 \le t$,
\bq
\label{max_speed}
v( t ) \le v\submax( t ) \equiv v_0 + \hat{ g } t.
\eq
We note that equality in \eqref{max_speed} is reached exactly when $t = 0$; for $t > 0$, the quantity $v\submax$ is the supremum of $v$ across all sets $\mathcal{ I }$ of initial conditions and parameters under consideration in this document, representing most closely the case where the projectile begins at rest (so that the force of a tailwind is maximal), very close to the surface of the Earth (so that the acceleration of gravity is well-approximated by $g$), with wind in the radially inward direction (so that the forces of gravity and wind directly add).  Integrating \eqref{max_speed}, we find that the projectile's distance $| \vec{ r } |$ from its initial position satisfies, for $0 \le t$,
\bq
\label{max_distance}
| \vec{ r }( t ) | \le r\submax( t ) \equiv v_0 t + \tfrac{ 1 }{ 2 } \hat{ g } t^2.
\eq
To find the value of $t$ for which $r\submax( t ) = r\submax^\star ~ \exists ~ r\submax^\star \ge 0$, we solve the quadratic equation
\bq
\tfrac{ 1 }{ 2 } \hat{ g } t^2 + v_0 t - r\submax^\star = 0,
\eq
whose only identically non-negative solution is
\bq
t_+ \equiv \frac{ \sqrt{ v_0^2 + 2 \hat{ g } r\submax^\star } - v_0 }{ \hat{ g } }.
\eq
Now, $t_+$ represents the time required for the projectile to reach the prescribed distance $r\submax^\star$ from its initial position, in the case where the projectile is moving in a straight line away from its initial position, at its maximum possible speed at every point in time.  Thus $t_+$ is the minimum possible time for the projectile to move the distance $r\submax^\star$ from its initial position.  In other words, $t_+$ is a tight upper bound for the maximum time for which we can guarantee -- across all sets $\mathcal{ I }$ of initial conditions and parameters under consideration in this document -- that the projectile has moved no farther than $r\submax^\star$ from its initial position:
\bq
t_+ = \sup_{ \mathcal{ I } }\{  0 \le t : | \vec{ r }( t ) | \le r\submax^\star \}.
\eq
With the condition \eqref{rho_valid_condition} in mind, we use the relationship \eqref{scaling_eqn} to set
\bq
r\submax^\star = r\c \rho\submax,
\eq
and we use the relationship \eqref{t_t_c_eqn} to divide $t_+$ by $t\c$ in order to find the scaled, dimensionless time
\begin{empheq}[box=\mybox]{equation}
\label{tau_max_safe}
\tau\submax\safe \equiv \frac{ \sqrt{ v_0^2 + 2 \hat{ g } r\c \rho\submax } - v_0 }{ t\c \hat{ g } }.
\end{empheq}
The quantity $\tau\submax\safe$ satisfies the conditions \eqref{rho_valid_condition}, \eqref{tau_valid_condition}, and \eqref{tau_desired_condition}; in fact, $\tau\submax\safe$ is the infimum of $\tau_\star$ across all sets $\mathcal{ I }$ of initial conditions and parameters under consideration in this document.  The quantity $\tau\submax\safe$ plays a crucial role in controlling the error in the computation of trajectories based on our results; see Algorithms \ref{alg:subtraj} and \ref{alg:complete_traj} for additional details.

\section{Analysis of zeroth-order trajectories with constant wind}\label{sec:const_wind_analysis}

Here we discuss in detail the $\O( 1 )$ solutions subject to the constant-wind condition \cw.

\subsection{Time of tangency}\label{sec:tan_time}

We now consider the problem of finding the value of $\tau$ at which the projectile's motion is tangential to the surface of the Earth.

\subsubsection{General condition for the time of tangency}

We denote by $\tau\tang$ the value of $\tau$ satisfying the tangency condition\footnote{We could also have used the condition $\absposv \cdot \rhovprime = 0$, which produces the same $\O( 1 )$ condition \eqref{eq:tan_tau_condn}.}
\bq
\label{eq:tan_tau_condn_general}
\abspos'( t ) = 0.
\eq
To $\O( 1 )$, the condition \eqref{eq:tan_tau_condn_general} is equivalent to
\bq
\label{eq:tan_tau_condn}
\av \cdot \rhovprime_0 = 0.
\eq
We note that we can always relate the value of $\tau\tang$ to the corresponding value $t\tang$ in the original time scale via the relationship \eqref{t_t_c_eqn}.

\subsubsection{Zeroth-order time of tangency for $b\eff = 0$}

Using the solution $\rhovprime_0( \tau ; \bz )$ of \eqref{rho_0_bz}, we find that the $\O( 1 )$ condition \eqref{eq:tan_tau_condn} becomes
\begin{subequations}
\label{tau_tan_bz_condn}
\begin{align}
\av \cdot( \dv + \gv \tau )	& = 0\\
\d \s + \g \tau 				& = 0,
\end{align}
\end{subequations}
whose solution in the variable $\tau$ is
\begin{empheq}[box=\mybox]{equation}
\label{tau_tan_bz}
\tau\tang[ \bz ] \equiv \frac{ \d \s }{ | \g | },
\end{empheq}
valid in the interval $[ 0 , \tau_\star ]$.  Here, the notation $\tau\tang[ \bz ]$ indicates that the quantity $\tau\tang$ has inherited the condition \bz{} from the solution $\rhovprime_0( \tau ; \bz )$ used here.

\paragraph{Physical interpretation} We point out that the RHS of \eqref{tau_tan_bz} is the ratio of the component of the initial scaled velocity $\dv$ in the initially radially outward direction $\av$ to the scaled acceleration of strength $| \g |$ induced by the zeroth-order force of gravity in the initially radially inward direction $-\av$.  That is, the scaled time $\tau\tang[ \bz ]$ is approximated by the time of ascent of a projectile problem along the $\av$-axis, in which the projectile has initial outward speed $\d \s$ and is subjected only to a constant inward gravitational acceleration of magnitude $| \g |$.  In addition, for the case $b\eff = 0$, the quantity $\tau\tang$ is the same as the quantity $\tau\ascent$, the time required for the projectile to reach the highest point of its trajectory.

\subsubsection{Zeroth-order time of tangency for $b\eff \ne 0$ and constant wind}

Using the solution $\rhovprime_0( \tau ; \bnz , \cw )$ of \eqref{rhoprime_0_bnz_cw}, we find that the $\O( 1 )$ condition \eqref{eq:tan_tau_condn} becomes
\bq
\label{linear_R_max_condition}
\av \cdot ( \Cv + \Dv \expmt ) = 0.
\eq
\paragraph{Case 1} For $\av \cdot \Cv = 0$, the outward force of wind acting on the projectile when it has no radial motion exactly cancels gravity's inward force, and the condition \eqref{linear_R_max_condition} becomes $\d \s \expmt = 0$.  For $\d \s \ne 0$ there are no solutions: the projectile continues in its initial direction having a non-tangential component, never reaching a point where it is tangential to the surface of the Earth.  For $\d \s = 0$, on the other hand, every $\tau$ is a solution, as the projectile is, to $\O( 1 )$, always tangential to the surface of the Earth: when $\d = 0$, the projectile is suspended in the air, motionless; when $\d \ne 0$ but $\s = 0$, the motion of the projectile is, to $\O( 1 )$, constrained to lie along the direction of its initial, tangential velocity.

\paragraph{Case 2} If $\av \cdot \Cv \ne 0$ and $\av \cdot \Dv = 0$, then there are no solutions.

\paragraph{Case 3} If $\av \cdot \Cv \ne 0$ and $\av \cdot \Dv \ne 0$, then we find the solution to the $\O( 1 )$ condition \eqref{linear_R_max_condition} in the variable $\tau$ to be
\begin{empheq}[box=\mybox]{equation}
\label{tau_ascent_linear}
\tau\tang[ \bnz , \cw ] \equiv \td \ln\left( -\frac{ \av \cdot \Dv }{ \av \cdot \Cv } \right),
\end{empheq}
valid in the interval $[ 0 , \tau_\star ]$.

\paragraph{Physical interpretation} We observe that the vector $-\Dv$ points from the projectile's initial velocity to its asymptotic velocity: $-\Dv = \rhovprime_0( \infty ; \bnz , \cw ) - \rhovprime_0( 0 ; \bnz , \cw )$.  On the other hand, the vector $\Cv$ is the asymptotic velocity of the projectile: $\Cv = \rhovprime_0( \infty ; \bnz , \cw )$.  Thus the argument of the logarithm of \eqref{tau_ascent_linear} is the ratio of the radially outward component of these two vectors.

\subsection{Flight time}\label{sec:flight_time}

We consider the motion of the projectile beginning at the prescribed initial conditions, continuing until the projectile reaches some prescribed final radius $R\f$ satisfying
\bq
\label{R_f_min}
R\f \ge R\E.
\eq
In cases for which there are both pre-tangent and post-tangent solutions, we provide both solutions.

\subsubsection{General condition for the flight time}\label{sec:flight_time}

The flight time can be estimated from the condition
\begin{subequations}
\label{R_f_general_condn}
\begin{align}
R\f
	& = \abspos\\
	& = R_0 \left[ 1 + \eps ( \av \cdot \rhov_0 ) + \O( \eps^2 ) \right],
\end{align}
\end{subequations}
where we have used \eqref{distance_expansion_eps}.  Taking into account terms up to $\O( 1 )$, we find the general condition
\bq
\label{flight_time_temp}
0 = \mu + ( \av \cdot \rhov_0 )
\eq
for the flight time, where
\bq
\label{mu_defn}
\mu \equiv \frac{ R_0 - R\f }{ r\c }.
\eq
We note that $\mu$ is a dimensionless length in $\rho$-space.  For this reason, the analyses in the subsections to follow are valid only when
\bq
\label{mu_valid_condn}
| \mu | \le \rho\submax.
\eq

\subsubsection{Zeroth-order flight time for $b\eff = 0$}\label{sec:vacuum_flight_time}

From the condition \eqref{flight_time_temp} and the solution $\rhov_0( \tau ; \bz )$ of \eqref{rho_0_bz}, we find the condition
\bq
\label{vacuum_flight_time_condition}
0 = \mu + \d \s \tau + \tfrac{ 1 }{ 2 } \g \tau^2,
\eq
whose solutions are
\bq
\label{flight_time}
\tau = \frac{ \d \s }{ | \g | } \pm \sqrt{ \psi },
\eq
where
\bq
\label{psi_defn}
\psi \equiv \frac{ \d^2 \s^2 }{ \g^2 } + \frac{ 2 \mu }{ | \g | }.
\eq
In order to select the inward ($\abspos' < 0$) solution corresponding to the later, post-tangent flight time, we define the estimated flight time to be
\begin{empheq}[box=\mybox]{equation}
\label{flight_time_vacuum}
\tau\flight\posttan[ \bz ] \equiv \frac{ \d \s }{ | \g | } + \sqrt{ \psi },
\end{empheq}
valid in the interval $[ 0 , \tau_\star ]$.  We note that $\tau\flight\posttan[ \bz ]$ is real and non-negative if and only if $\psi \ge 0$; this occurs if and only if $R\f \le R\submax[ \bz ]$, where $R\submax[ \bz ]$ is defined in \eqref{R_max_vacuum}.

We also define the outward ($\abspos' > 0$) solution corresponding to the earlier, pre-tangent flight time to be
\begin{empheq}[box=\mybox]{equation}
\label{flight_time_outward}
\tau\flight\pretan[ \bz ] \equiv \frac{ \d \s }{ | \g | } - \sqrt{ \psi },
\end{empheq}
valid in the interval $[ 0 , \tau_\star ]$.  We note that $\tau\flight\pretan[ \bz ]$ is real if and only if $\psi \ge 0$, which occurs if and only if $R\f \le R\submax[ \bz ]$.  Furthermore, when $\psi \ge 0$, we find that $\tau\flight\pretan[ \bz ] \ge 0$ if and only if both $\s \ge 0$ and $R_0 \le R\f$.

\paragraph{Physical interpretation} We can write the solutions in $\tau$ as
\bq
\tau = \tau\tang[ \bz ] \left( 1 \pm \sqrt{ 1 + \frac{ \mu }{ \Delta\rho } } \right),
\eq
where
\bq
\label{defn:Delta_rho}
\Delta\rho \equiv \frac{ \d^2 \s^2 }{ 2 | \g | }
\eq
is the maximum radial displacement of the projectile in $\rho$-space (see Section \ref{sec:R_max_vacuum}).

\subsubsection{Zeroth-order flight time for $b\eff \ne 0$ and constant wind}\label{sec:Lambert_discussion}

From the condition \eqref{flight_time_temp} and the solution $\rhov_0( \tau ; \bnz , \cw )$ of \eqref{rho_0_bnz_cw}, we find that
\begin{subequations}
\label{linear_flight_time_condition}
\begin{align}
0
	& = \mu + ( \av \cdot \rhov_0 )\\
	& = \mu + ( \av \cdot \Cv ) \tau - \td ( \av \cdot \Dv ) ( \expmt - 1 )\\
	& = a_1 + a_2 \tau + a_3 \expmt,
\end{align}
\end{subequations}
where
\begin{subequations}
\begin{align}
a_2 & \equiv ( \av \cdot \Cv ),\\
a_3 & \equiv -\td ( \av \cdot \Dv ),\\
a_1 & \equiv \mu - a_3.
\end{align}
\end{subequations}
\paragraph{Case 1} If $\av \cdot \Cv = 0$ and $\d \s = 0$, then \eqref{linear_flight_time_condition} has a solution if and only if $\mu = 0$, in which case every $\tau$ is a solution.  In this case, $R_0 = R\f$ and the projectile is initially at radial rest (either $\d = 0$ and the projectile is at rest, or $\s = 0$ and the projectile's motion is initially entirely tangential) with no net radial force, implying that, to $\O( 1 )$, it remains at radius $R\f$.

\paragraph{Case 2} If $\av \cdot \Cv = 0$ and $\d \s \ne 0$, then the solution to \eqref{linear_flight_time_condition} is given by
\bq
\tau\flight[ \av \cdot \Cv = 0 , \d \s \ne 0 ] = -\td \ln\left( 1 + \frac{ \mu b\eff }{ \d \s } \right).
\eq

\paragraph{Case 3} If $\av \cdot \Cv \ne 0$ and $\av \cdot \Dv = 0$, then the solution to \eqref{linear_flight_time_condition} is given by
\bq
\tau\flight[ \av \cdot \Cv \ne 0 , \av \cdot \Dv = 0 ] = -\frac{ \mu }{ \av \cdot \Cv }.
\eq

\paragraph{Case 4} If $\av \cdot \Cv \ne 0$ and $\av \cdot \Dv \ne 0$, then each solution to \eqref{linear_flight_time_condition} may be written as
\bq
\label{linear_flight_time_unspecified}
\tau = \td W\left( W\subarg \right) - \frac{ a_1 }{ a_2 },
\eq
where
\bq
\label{W_arg_defn}
W\subarg \equiv -\frac{ a_3 b\eff }{ a_2 } \exp\left( \frac{ a_1 b\eff }{ a_2 } \right)
\eq
and where $W$ is the Lambert W function (also called the product log function).

We focus on two branches of the Lambert W function \cite{nist}, each of which we view as a function $f : \mathbb{ R } \to \mathbb{ R }$.  The principal branch, denoted $W_0$, is an increasing function with domain $[ -1 / e , \infty )$ and range $[ -1 , \infty )$; the branch denoted $W_{ -1 }$ is a decreasing function with domain $[ -1 / e , 0 )$ and range $( -\infty , -1 ]$.  For $-1 / e < x < 0$, these branches satisfy $W_{ -1 }( x ) < W_0( x )$, with $W_{ -1 }( -1 / e ) = W_0( -1 / e ) = -1$.  We also have that $W_{ -1 }( x e^x ) = x$ for $x \le -1$ and $W_0( x e^x ) = x$ for $x \ge -1$.

\paragraph{Physical interpretation of the branches of the Lambert W function} We physically interpret these branches of the multi-valued function $W$ in analogy with the branches of the multi-valued function $\sqrt{ \, \cdot \, }$.  For the function $\sqrt{ \, \cdot \, }$, we select the branch $+\sqrt{ \, \cdot \, }$ when we wish to study the later, post-tangent flight time, as in \eqref{flight_time_vacuum}, and we select the branch $-\sqrt{ \, \cdot \, }$ when we wish to study the earlier, pre-tangent flight time, as in \eqref{flight_time_outward}.  The same reasoning can be applied to the function $W$: one branch characterizes the part of the trajectory occurring before the projectile reaches tangency, while the other branch characterizes the part of the trajectory occurring after the projectile reaches tangency (with the branches converging at the point of tangency, where the trajectory realizes its extreme\footnote{In the case of $b\eff \ne 0$, we discuss \textsl{extreme} radii -- rather than a \textsl{maximum} radius -- because the extreme radius can be either a maximum \textsl{or} a minimum (with the help of a radially outward wind force); see Figure \ref{fig:trajectory_types} for schematic representations of such trajectories, and see Figure \ref{fig:raindrop} for a numerical example of a trajectory reaching an extreme radius that is a minimum.} radius).

In order to select the post-tangent solution corresponding to the later flight time, we define the estimated flight time to be
\begin{empheq}[box=\mybox]{equation}
\label{flight_time_linear}
\tau\flight\posttan[ \bnz , \cw ] \equiv \td W_0\left( W\subarg \right) - \frac{ a_1 }{ a_2 },
\end{empheq}
valid in the interval $[ 0 , \tau_\star ]$.

We also define the pre-tangent solution corresponding to the earlier flight time as
\begin{empheq}[box=\mybox]{equation}
\label{tau_pretan_bnz}
\tau\flight\pretan[ \bnz , \cw ] \equiv \td W_{ -1 }\left( W\subarg \right) - \frac{ a_1 }{ a_2 },
\end{empheq}
valid in the interval $[ 0 , \tau_\star ]$.

\subsection{Extreme radii}

Here we investigate the extreme values of the radius achieved by the projectile when its trajectory is not interrupted by the Earth and is not terminated by having reached the final radius $R\f$.  In the case \bz, this is always a maximum radius; in the case \bnz, this can be either a maximum or a minimum radius, the latter possible in the case of wind having a radially outward component -- see Figure \ref{fig:trajectory_types} for schematic representations of some of these cases, and see Figure \ref{fig:raindrop} for a numerical example.

\newlength{\subfigwidth}
\setlength{\subfigwidth}{1.25in}

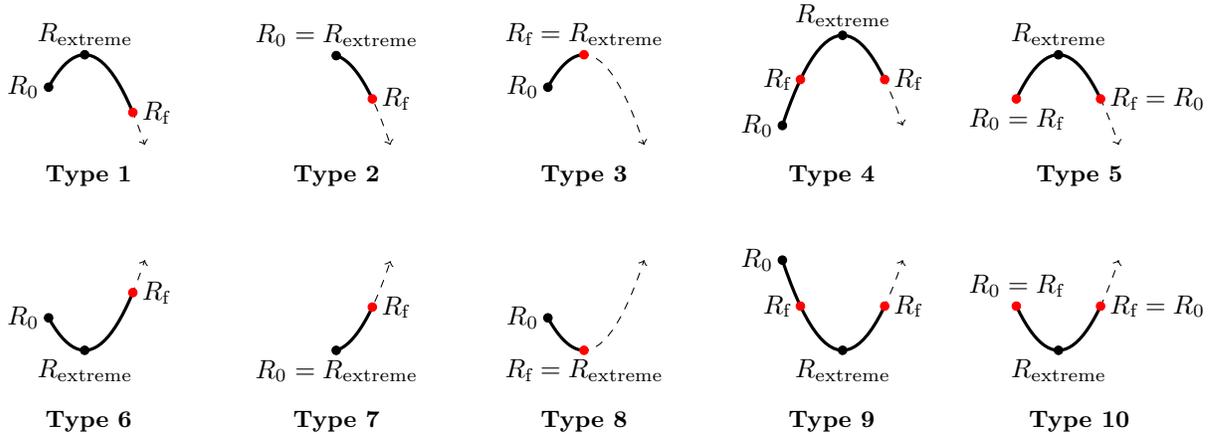
\begin{figure}[H]
	\captionsetup[subfigure]{justification=centering,labelformat=empty}
	\centering
	\begin{subfigure}[t]{\subfigwidth}
	\centering
	\begin{tikzpicture}[scale=0.8]	
		\draw[->, scale = 1, domain = -0.6:1, variable = \x, dashed]  plot ({\x},{-\x*\x*1.5});
		\draw[scale = 1, domain = -0.6:0.8, variable = \x, very thick]  plot ({\x},{-\x*\x*1.5});
		\draw[red,fill=red] (0.8,-0.96) circle[radius=2pt];
		\node[right] at (0.8,-0.96) {$R\f$};
		\draw[fill] (-0.6, -0.54) circle[radius=2pt];
		\node[left] at (-0.6,-0.54) {$R_0$};
		\draw[fill] (0,0) circle[radius=2pt];
		\node[above] at (0,0) {$R\subext$};
	\end{tikzpicture}
	\caption{\textbf{Type 1}}
	\end{subfigure}
	\begin{subfigure}[t]{\subfigwidth}
	\centering
	\begin{tikzpicture}[scale=0.8]	
		\draw[->, scale = 1, domain = 0.1:1, variable = \x, dashed]  plot ({\x},{-\x*\x*1.5});
		\draw[scale = 1, domain = 0.1:0.7, variable = \x, very thick]  plot ({\x},{-\x*\x*1.5});
		\draw[fill] (0.1, -0.015) circle[radius=2pt];
		\node[above] at (0.1,-0.015) {$R_0 = R\subext$};
		\draw[red,fill=red] (0.7, -0.735) circle[radius=2pt];
		\node[right] at (0.7,-0.735) {$R\f$};
	\end{tikzpicture}
	\caption{\textbf{Type 2}}
	\end{subfigure}
	\begin{subfigure}[t]{\subfigwidth}
	\centering
	\begin{tikzpicture}[scale=0.8]	
		\draw[->, scale = 1, domain = -0.6:1, variable = \x, dashed]  plot ({\x},{-\x*\x*1.5});
		\draw[scale = 1, domain = -0.6:0, variable = \x, very thick]  plot ({\x},{-\x*\x*1.5});
		\draw[fill] (-0.6,-0.54) circle[radius=2pt];
		\node[left] at (-0.6,-0.54) {$R_0$};
		\draw[red,fill=red] (0,0) circle[radius=2pt];
		\node[above] at (0,0) {$R\f = R\subext$};
	\end{tikzpicture}
	\caption{\textbf{Type 3}}
	\end{subfigure}
	\begin{subfigure}[t]{\subfigwidth}
	\centering
	\begin{tikzpicture}[scale=0.8]	
		\draw[->, scale = 1, domain = -1:1, variable = \x, dashed]  plot ({\x},{-\x*\x*1.5});
		\draw[scale = 1, domain = -1:0.7, variable = \x, very thick]  plot ({\x},{-\x*\x*1.5});
		\draw[fill] (-1,-1.5) circle[radius=2pt];
		\node[left] at (-1,-1.5) {$R_0$};
		\draw[red,fill=red] (0.7,-0.735) circle[radius=2pt];
		\node[right] at (0.7,-0.735) {$R\f$};
		\draw[red,fill=red] (-0.7, -0.735) circle[radius=2pt];
		\node[left] at (-0.7,-0.735) {$R\f$};
		\draw[fill] (0,0) circle[radius=2pt];
		\node[above] at (0,0) {$R\subext$};
	\end{tikzpicture}
	\caption{\textbf{Type 4}}
	\end{subfigure}
	\begin{subfigure}[t]{\subfigwidth}
	\centering
	\begin{tikzpicture}[scale=0.8]	
		\draw[->, scale = 1, domain = -0.7:1, variable = \x, dashed]  plot ({\x},{-\x*\x*1.5});
		\draw[scale = 1, domain = -0.7:0.7, variable = \x, very thick]  plot ({\x},{-\x*\x*1.5});
		\draw[red,fill=red] (0.7,-0.735) circle[radius=2pt];
		\node[right] at (0.7,-0.735) {$R\f = R_0$};
		\draw[red,fill=red] (-0.7, -0.735) circle[radius=2pt];
		\node[below] at (-0.7,-0.735) {$R_0 = R\f$};
		\draw[fill] (0,0) circle[radius=2pt];
		\node[above] at (0,0) {$R\subext$};
	\end{tikzpicture}
	\caption{\textbf{Type 5}}
	\end{subfigure}	
	
	\medskip\medskip\medskip
	
	\begin{subfigure}[t]{\subfigwidth}
	\centering
	\begin{tikzpicture}[scale=0.8]	
		\draw[->, scale = 1, domain = -0.6:1, variable = \x, dashed]  plot ({\x},{\x*\x*1.5});
		\draw[scale = 1, domain = -0.6:0.8, variable = \x, very thick]  plot ({\x},{\x*\x*1.5});
		\draw[red,fill=red] (0.8,0.96) circle[radius=2pt];
		\node[right] at (0.8,0.96) {$R\f$};
		\draw[fill] (-0.6, 0.54) circle[radius=2pt];
		\node[left] at (-0.6,0.54) {$R_0$};
		\draw[fill] (0,0) circle[radius=2pt];
		\node[below] at (0,0) {$R\subext$};
	\end{tikzpicture}
	\caption{\textbf{Type 6}}
	\end{subfigure}
	\begin{subfigure}[t]{\subfigwidth}
	\centering
	\begin{tikzpicture}[scale=0.8]	
		\draw[->, scale = 1, domain = 0.1:1, variable = \x, dashed]  plot ({\x},{\x*\x*1.5});
		\draw[scale = 1, domain = 0.1:0.7, variable = \x, very thick]  plot ({\x},{\x*\x*1.5});
		\draw[fill] (0.1, 0.015) circle[radius=2pt];
		\node[below] at (0.1,0.015) {$R_0 = R\subext$};
		\draw[red,fill=red] (0.7, 0.735) circle[radius=2pt];
		\node[right] at (0.7,0.735) {$R\f$};
	\end{tikzpicture}
	\caption{\textbf{Type 7}}
	\end{subfigure}
	\begin{subfigure}[t]{\subfigwidth}
	\centering
	\begin{tikzpicture}[scale=0.8]	
		\draw[->, scale = 1, domain = -0.6:1, variable = \x, dashed]  plot ({\x},{\x*\x*1.5});
		\draw[scale = 1, domain = -0.6:0, variable = \x, very thick]  plot ({\x},{\x*\x*1.5});
		\draw[fill] (-0.6,0.54) circle[radius=2pt];
		\node[left] at (-0.6,0.54) {$R_0$};
		\draw[red,fill=red] (0,0) circle[radius=2pt];
		\node[below] at (0,0) {$R\f = R\subext$};
	\end{tikzpicture}
	\caption{\textbf{Type 8}}
	\end{subfigure}
	\begin{subfigure}[t]{\subfigwidth}
	\centering
	\begin{tikzpicture}[scale=0.8]	
		\draw[->, scale = 1, domain = -1:1, variable = \x, dashed]  plot ({\x},{\x*\x*1.5});
		\draw[scale = 1, domain = -1:0.7, variable = \x, very thick]  plot ({\x},{\x*\x*1.5});
		\draw[fill] (-1,1.5) circle[radius=2pt];
		\node[left] at (-1,1.5) {$R_0$};
		\draw[red,fill=red] (0.7,0.735) circle[radius=2pt];
		\node[right] at (0.7,0.735) {$R\f$};
		\draw[red,fill=red] (-0.7, 0.735) circle[radius=2pt];
		\node[left] at (-0.7,0.735) {$R\f$};
		\draw[fill] (0,0) circle[radius=2pt];
		\node[below] at (0,0) {$R\subext$};
	\end{tikzpicture}
	\caption{\textbf{Type 9}}
	\end{subfigure}
	\begin{subfigure}[t]{\subfigwidth}
	\centering
	\begin{tikzpicture}[scale=0.8]	
		\draw[->, scale = 1, domain = -0.7:1, variable = \x, dashed]  plot ({\x},{\x*\x*1.5});
		\draw[scale = 1, domain = -0.7:0.7, variable = \x, very thick]  plot ({\x},{\x*\x*1.5});
		\draw[red,fill=red] (0.7,0.735) circle[radius=2pt];
		\node[right] at (0.7,0.735) {$R\f = R_0$};
		\draw[red,fill=red] (-0.7, 0.735) circle[radius=2pt];
		\node[above] at (-0.7,0.735) {$R_0 = R\f$};
		\draw[fill] (0,0) circle[radius=2pt];
		\node[below] at (0,0) {$R\subext$};
	\end{tikzpicture}
	\caption{\textbf{Type 10}}
	\end{subfigure}	
\caption{Schematic representations of some possible trajectory types, according to the relative values of $R_0$, $R\f$, and $R\subext$.  Heavy lines are realized trajectories.  Dashed arrows indicate the direction of movement.  Red dots are solutions to $\abspos = R\f$.  See Figures \ref{fig:point_mass} -- \ref{fig:beach_ball} for numerical examples of some of these trajectory types.}
\label{fig:trajectory_types}
\end{figure}

\subsubsection{Maximum radius for $b\eff = 0$}\label{sec:R_max_vacuum}

In the case \bz, we found the flight time by choosing the appropriate branch of the multi-valued function $\sqrt{ \, \cdot \, }$, viewing each branch as a function $f : \mathbb{ R } \to \mathbb{ R }$.  The branch with greater values corresponded to the later flight time (the inward solution), while the branch with lesser values corresponded to the earlier flight time (the outward solution).  Viewing these branches as functions of the desired final radius $R\f$, these branches converge to a common value when $R\f = R\submax$ (that is, when the earlier and later flight times coincide, at the radial peak of the trajectory).  For the case \bz, this is when
\bq
\label{R_max_condition_vacuum}
\psi = 0.
\eq
Solving the condition \eqref{R_max_condition_vacuum} for $R\f$ and using this value as our approximation of $R\submax$, we define
\begin{empheq}[box=\mybox]{equation}
\label{R_max_vacuum}
R\submax[ \bz ] \equiv R_0 \left( 1 + \eps \frac{ \d^2 \s^2 }{ 2 | \g | } \right),
\end{empheq}
valid when $\tau\tang[ \bz ]$ is valid and when the condition \eqref{mu_valid_condn} holds.  We note that $R\submax[ \bz ] \ge R_0$.

\paragraph{Physical interpretation} We point out the similarity between the fractional term of our result and the maximum height of an elementary projectile in a vacuum: $\d \s$ is the dimensionless initial speed in the radially outward direction, and $| \g |$ is the local dimensionless strength of gravity.  We also recognize the quantity multiplying $\eps$ to be $\Delta\rho$ of \eqref{defn:Delta_rho}.

\subsubsection{Extreme radius for $b\eff \ne 0$ and constant wind}\label{sec:R_max_linear}

Following the same argument for the case of non-negligible air resistance and constant wind -- but replacing the multi-valued function $\sqrt{ \, \cdot \, }$ with the multi-valued function $W$ -- we find that the earlier and later flight times coincide when
\bq
\label{R_max_condition_linear}
W\subarg = -\exp( -1 ).
\eq
Solving the condition \eqref{R_max_condition_linear} for $R\f$ and using this value as our approximation of $R\subext$, we define
\begin{empheq}[box=\mybox]{equation}
\label{R_max_linear}
R\subext[ \bnz , \cw ] \equiv R_0 \left( 1 + \eps \td a_2 \left[ 1 - \frac{ a_3 b\eff }{ a_2 } + \ln\left( \frac{ a_3 b\eff }{ a_2 } \right) \right] \right),
\end{empheq}
valid when $\tau\tang[ \bnz , \cw ]$ is valid and when the condition \eqref{mu_valid_condn} holds.

\paragraph{Physical interpretation} The quantity multiplying $\eps$ in \eqref{R_max_linear} can be written as
\bq
[ \av \cdot \rhovprime_0( 0 ; \bnz , \cw ) ] \td + [ \av \cdot \rhovprime_0( \infty ; \bnz , \cw ) ] \tau\tang[ \bnz , \cw ];
\eq
from this, we see that the initial velocity of the projectile effectively contributes to a change in its radial distance over its time scale of decay, $\td$, and that the asymptotic velocity of the projectile effectively contributes to a change in its radial distance over the entire time scale $\tau\tang$ over which the projectile reaches its extreme radius.

\subsection{A note on the consistency of our results}\label{sec:consistency}

As a check on the consistency of our results, each of the following can be shown.
\begin{enumerate}[label={(\textbf{\arabic*})}]

	\item Pre-tangent and post-tangent flight times both converge to the time of tangency when the condition for branch equality is met:
	\begin{subequations}
	\bq
	\left.\tau\flight\pretan[ \bz ]\right|_{ \eqref{R_max_condition_vacuum} } = \left.\tau\flight\posttan[ \bz ]\right|_{ \eqref{R_max_condition_vacuum} } = \tau\tang[ \bz ]
	\eq
	and
	\bq
	\left.\tau\flight\pretan[ \bnz , \cw ]\right|_{ \eqref{R_max_condition_linear} } = \left.\tau\flight\posttan[ \bnz , \cw ]\right|_{ \eqref{R_max_condition_linear} } = \tau\tang[ \bnz , \cw ].
	\eq
	\end{subequations}
	
	\item To $\O( 1 )$, extreme radii can be found by computing the distance from the projectile to the center of the Earth at the time of tangency:
	\begin{subequations}
	\bq
	| \absposv_0 + r\c \rhov_0( \tau\tang[ \bz ] ; \bz ) | \sim R\submax[ \bz ]
	\eq
	and
	\bq
	| \absposv_0 + r\c \rhov_0( \tau\tang[ \bnz , \cw ] ; \bnz , \cw ) | \sim R\subext[ \bnz , \cw ].
	\eq
	\end{subequations}

\end{enumerate}

\section{Some numerical considerations}

Here we discuss some considerations relevant to our numerical implementation of our solutions.

\subsection{Estimating the parameter $b\drag$}\label{sec:choosing_b_drag}

From \cite[p.~44]{Taylor}, the linear drag coefficient $b\drag$ may be written, in the case of a spherical projectile of diameter $D$, as
\bq
\label{b_c_vals}
b\drag = \z_b D,
\eq
where $\z_b$ is approximated, in the case of air at standard temperature and pressure, by \cite[p.~44]{Taylor}
\bq
\z_b \approx 1.6 \times 10^{ -4 } \text{ Newtons $\cdot$ seconds / meters$^2$}.
\eq
We note that the value of $\z_b$ is dependent upon the properties of the medium: in order to ensure accuracy, one must consider the temperature and pressure of the medium, as well as its fluid characteristics at the speeds over which one wishes to compute projectile motion.

\subsection{Small values of $b\eff$}\label{sec:b_eff_small}

We have provided solutions for both $b\eff = 0$ and $b\eff \ne 0$, but there is a third regime that is important in numerical implementations, where $0 < b\eff \ll 1$.  In this regime, $\td \gg 1$.  Since the solutions for $b\eff \ne 0$ involve powers of $\td$, these solutions can become numerically unstable.  For this reason, in Appendix \ref{sec:b_eff_expansions}, we have provided expanded solutions from the application of a perturbation method with small parameter $b\eff$.  This third regime for $b\eff$ is taken into account in our numerical implementations of Algorithms \ref{alg:subtraj} and \ref{alg:complete_traj}; see the Matlab code \cite{matlab-code} for complete details.

\subsection{Parameterized, error-controlled algorithms}\label{sec:algorithms}

In Section \ref{sec:model_validity}, we showed that our models would be valid within the time interval $[ 0 , \tau\submax\safe ]$.  When numerically computing a trajectory, then, we restrict our solutions to this time interval; Algorithm \ref{alg:subtraj} provides one method of doing this.  If the trajectory has not yet ended -- via intersection with either the Earth or with the final radius $R\f$ -- by the time $\tau\submax\safe$ occurs, then we evaluate the positions and velocities at $\tau\submax\safe$ and feed them back into the calculations as the new initial conditions.  Algorithm \ref{alg:complete_traj} illustrates this process of piecing together subtrajectories to form a complete trajectory and is implemented in the Matlab code \cite{matlab-code}.  We note that, in our implementation, we ignore any pre-tangent intersection with $R\f$, choosing to continue the trajectory until either a post-tangent intersection with $R\f$ or an intersection with the Earth.  In addition, we have chosen to implement only the constant-wind solutions of Section \ref{sec:const_wind} and Appendix \ref{sec:b_eff_expansions}.
\begin{algorithm}[H]
\centering
\begin{pseudo}[fullwidth, line-height=1.25]*
\toprule
	\multicolumn{3}{c}{\textsc{Subtrajectory}}
	\\[bol=\midrule]
	Compute $\tau\submax\safe$
	& See \eqref{tau_max_safe}
	\\
	$( \tau\flight\posttan , \tau\E\pretan , \tau\E\posttan ) \leftarrow ( \infty , \infty , \infty )$
	& Initialize
	\\
	Compute $\mu$
	& See \eqref{mu_defn}
	\\
	\kw{if} $| \mu | \le \rho\submax$
	& See \eqref{mu_valid_condn}
	\\+
		Compute $\tau\flight\posttan$
		& See \eqref{flight_time_vacuum}, \eqref{flight_time_linear}, and \eqref{tau_bs}
		\\-
	\kw{end}
	&
	\\
	Compute $\mu\E \equiv \mu|_{ R\f = R\E }$
	& 
	\\
	\kw{if} $| \mu\E | \le \rho\submax$
	& See \eqref{mu_valid_condn}
	\\+
		Compute $\tau\E\pretan \equiv \tau\flight\pretan|_{ R\f = R\E }$
		& See \eqref{flight_time_outward}, \eqref{tau_pretan_bnz}, and \eqref{tau_bs}
		\\
		Compute $\tau\E\posttan \equiv \tau\flight\posttan|_{ R\f = R\E }$
		& See \eqref{flight_time_vacuum}, \eqref{flight_time_linear}, and \eqref{tau_bs}
		\\-
	\kw{end}
	&
	\\
	$\tau\final \leftarrow \min\{ \tau\submax\safe , \tau\flight\posttan , \tau\E\pretan , \tau\E\posttan \}$
	& Effective final scaled time
	\\
	$\txt{isLastSubtrajectory} \leftarrow \bit{ \tau\final < \tau\submax\safe }$
	& Subtrajectory status
	\\
	Evaluate positions and velocities up to $t\final \equiv t\c \tau\final$
	& See Sections \ref{sec:solns_to_original} and \ref{sec:b_eff_expansion_solns}
	\\
	\kw{return} positions, velocities, times, and \txt{isLastSubtrajectory}
	&
	\\*
\bottomrule
\end{pseudo}
\caption{Evaluating a subtrajectory.}
\label{alg:subtraj}
\end{algorithm}
We note that Algorithm \ref{alg:complete_traj} requires only one evaluation of the vector-valued position and velocity per subtrajectory.  Using a root-finding procedure to estimate $\tau_\star$ (instead of using $\tau\submax\safe$, as we have done in Algorithm \ref{alg:subtraj}) may result in multiple such function evaluations per subtrajectory, but may also reduce the total number of required subtrajectories.  For performance-critical applications, we recommend further study of this trade-off, as we have made no attempt to optimize our numerical implementations of Algorithms \ref{alg:subtraj} and \ref{alg:complete_traj}.
\begin{algorithm}[H]
\centering
\begin{pseudo}[fullwidth, line-height=1.25]*
\toprule
	\multicolumn{3}{c}{\textsc{Complete Trajectory}}
	\\[bol=\midrule]
	$\txt{isLastSubtrajectory} \leftarrow \txt{false}$
	& Initialize
	\\
	\kw{while} \kw{not} \txt{isLastSubtrajectory}
	& Loop over subtrajectories
	\\+
		Compute subtrajectory and collect outputs
		& See Algorithm \ref{alg:subtraj}
		\\
		Update initial conditions for next subtrajectory
		& See the code \cite{matlab-code}
		\\-
	\kw{end}
	&
	\\
	\kw{return} collected positions, velocities, and times
	&
	\\*
\bottomrule
\end{pseudo}
\caption{Evaluating a complete trajectory.}
\label{alg:complete_traj}
\end{algorithm}

\section{Numerical results}\label{sec:numerical}

Here we present some numerical results.  All of our results were produced from the Matlab code \cite{matlab-code}, which we have made freely available on Matlab File Exchange.  The code therein implements Algorithms \ref{alg:subtraj} and \ref{alg:complete_traj}, outputs figures in the Matlab Live Script \txt{lpd\_\_demo.mlx}, and saves the data in the \txt{.csv} files that we used in generating Figures \ref{fig:point_mass} -- \ref{fig:beach_ball} of this document.

\subsection{Quantities of interest}\label{sec:traj_quantities}

We define the following quantities.  Here, $t_i$ is the $i\suptext{th}$ time step, with $0 \le i$ and $t_0 = 0$.  Quantities computed via our implementation of the methods of this document are denoted by $( \, \cdot \, )\subours$, and quantities computed via Matlab's numerical ODE solver are denoted by $( \, \cdot \, )\subnum$.
\begin{itemize}
	
	\item The relative error in the satisfaction of the ODE \eqref{original_model} is defined by the difference between the vector-valued LHS and vector-valued RHS of the original model \eqref{original_model}, divided by\footnote{We have chosen the RHS -- as opposed to the LHS -- as the denominator for no particular reason.} its RHS:
	\bq
	E\subODE( t_i ) \equiv \frac{ | \vec{ \eqref{original_model} }\subtext{LHS}( t_i ) - \vec{ \eqref{original_model} }\subtext{RHS}( t_i ) | }{ | \vec{ \eqref{original_model} }\subtext{RHS}( t_i ) | }.
	\eq

	\item The deviation in position is defined by
	\bq
	\Delta_{ \absposv }( t_i ) \equiv \left| \absposv\subours( t_i ) - \absposv\subnum( t_i ) \right|,
	\eq
	having units of meters.
	
	\item The deviation in velocity is defined by
	\bq
	\Delta_{ \absposvprime }( t_i ) \equiv \left| \absposvprime\subours( t_i ) - \absposvprime\subnum( t_i ) \right|,
	\eq
	having units of meters / second.

\end{itemize}
We call special attention to the fact that $E\subODE$ \textsl{is the key metric quantifying the performance of each method}.  This is because each method is designed to approximately solve the original ODE \eqref{original_model}.  In particular, the expanded solutions we provide are designed to improve the accuracy of the zeroth-order solutions \textsl{with respect to the original ODE \eqref{original_model}}.  This is \textsl{not} necessarily the same thing as improving the accuracy of the zeroth-order positions or velocities.  Indeed, in Figure \ref{fig:baseball}, we provide an example for which the expanded and numerical positions have the lowest values of $E\subODE$, whereas the zeroth-order and numerical positions are closer to one another.  Thus we focus on position and velocity \textsl{deviations} from the numerical solutions -- rather than \textsl{errors} -- because there is no guarantee that the numerical solutions more accurately compute the position and velocity than our solutions do, even in cases where the numerical solution achieves a lower value of $E\subODE$.

\subsection{Trajectory comparisons}\label{sec:traj_plots}

Below we present comparisons of trajectories and related quantities defined in Section \ref{sec:traj_quantities}, computed both via the methods of this document and via Matlab's \txt{ode15s} stiff ODE solver \cite{ode15s}, with absolute and relative tolerances set to \txt{5e-14}.

All trajectories we present have the following in common.
\begin{itemize}

	\item We have set the atmospheric thinning function to be $f\atm( s ) = e^{ -s }$, with characteristic length scale $\ell = 10,000$ meters.
	
	\item We have set the value of the parameter $b\eff^\star$ to be $b\eff^\star = \txt{1e-6}$ (empirically determined); see Appendix \ref{sec:b_eff_expansions} for more information about the parameter $b\eff^\star$.
	
	\item We have set the value of the parameter $\rho\submax$ to be $\rho\submax = 1$.
	
	\item In addition, the values of the error-control parameters $\eps\submax$, $\nu\submax$, and $\eta\submax$ were always set to be equal to one another; we call the common value of these error-control parameters $e\submax$, and we vary $e\submax$ -- and thus simultaneously vary $\eps\submax$, $\nu\submax$, and $\eta\submax$ together -- in the experiments to follow.\footnote{We have found empirically that the best accuracy is achieved when $e\submax \in [ \txt{5e-4} , \txt{1e-1} ]$, with smaller values generally resulting in better accuracy but more computational time.}
\end{itemize}
Table \ref{tab:fig_inputs} outlines some of the inputs used in producing Figures \ref{fig:point_mass} -- \ref{fig:beach_ball}.  We have tried to explore a reasonable variety of input parameter ranges and trajectory types\footnote{See the captions of Figures \ref{fig:point_mass} -- \ref{fig:beach_ball} for how each trajectory corresponds to a specific trajectory type from Figure \ref{fig:trajectory_types}.} without providing an overwhelming number of examples.  Details of the size\footnote{The size of the projectile is used in estimating the parameter $b\drag$, as in Section \ref{sec:choosing_b_drag}.} and mass of each projectile can be found in our numerical implementation \cite{matlab-code}.
\begin{table}[h!]
\begin{center}
\begin{tabular}{ccccccccccc}
\toprule
\addlinespace[0.5em]
Figure 					& Object		& $e\submax$	& $R_0$			& $v_0$		& $\th_r$				& $\phi_r$				& $\th_v$				& $\phi_v$				& $\vwind$			& $R\f$ \\
\addlinespace[0.5em]
\cmidrule(lr){1-11}
\addlinespace[0.5em]
\ref{fig:point_mass}	& point mass	& \txt{1e-3}	& $R\E$			& \txt{1e4}	& $\tfrac{ 1 }{ 3 }$	& $\tfrac{ 1 }{ 2 }$	& $\tfrac{ 1 }{ 10 }$	& $\tfrac{ 1 }{ 2 }$	& 									& $1.2 R\E$ \\
\addlinespace[0.5em]
\ref{fig:raindrop}		& raindrop		& \txt{1e-3}	& $R\E + 100$	& $50$		& 						& 						& $1$					& 						& $( 0 , 0 , \txt{5e5} )$	& $R\E + 80$ \\
\addlinespace[0.5em]
\ref{fig:golf_ball}		& golf ball		& \txt{5e-3}	& $R\E$			& $70$		&						& 						& $\tfrac{ 1 }{ 3 }$	& $\tfrac{ 1 }{ 2 }$	& $( -10 , 10 , 2 )$				& $R\E$ \\
\addlinespace[0.5em]
\ref{fig:baseball}		& baseball		& \txt{1e-2}	& $R\E$			& $30$		&						& $\tfrac{ 1 }{ 2 }$	& $\tfrac{ 1 }{ 4 }$	& $\tfrac{ 1 }{ 2 }$	& 					& $R\E$ \\
\addlinespace[0.5em]
\ref{fig:beach_ball}	& beach ball	& \txt{5e-4}	& $1.1 R\E$		& 			&						& 						&						& 						& $( 100 , -100 , -100 )$		& $1.05 R\E$ \\
\addlinespace[0.5em]
\bottomrule
\end{tabular}
\caption{Inputs for Figures \ref{fig:point_mass} -- \ref{fig:beach_ball}.  Omitted values are zeros.  Units are as follows: $e\submax$ (dimensionless); $R_0$ and $R\f$ (meters); $v_0$ and $\vwind$ (meters / second); $\th_r$, $\phi_r$, $\th_v$, and $\phi_v$ ($\pi$ radians).}
\label{tab:fig_inputs}
\end{center}
\end{table}

In Figures \ref{fig:point_mass} -- \ref{fig:beach_ball}, the time $t$ along each horizontal axis is expressed in units of seconds.  The radial trajectories show the distance $\abspos$ of the projectile from the center of the Earth, minus $R\E$, expressed in units of meters.

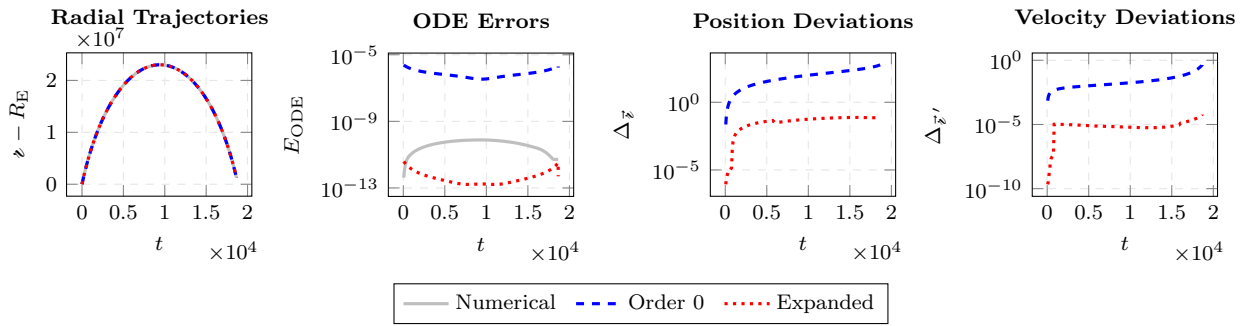
\addMatlabQuadFigureSmall{lpd__data__point_mass.csv}{Data for a point mass (generated with inputs from Table \ref{tab:fig_inputs}).  Since a point mass has size zero, it experiences no air resistance; as an additional quality check, the total energy of the projectile at the end of its trajectory is conserved to within the following fraction of the total energy of the projectile at the start of its trajectory: Numerical (\txt{6.0e-13}), Order 0 (\txt{9.0e-7}), Expanded (\txt{1.6e-11}).  This trajectory is an example of trajectory Type 4 of Figure \ref{fig:trajectory_types}.}{fig:point_mass}

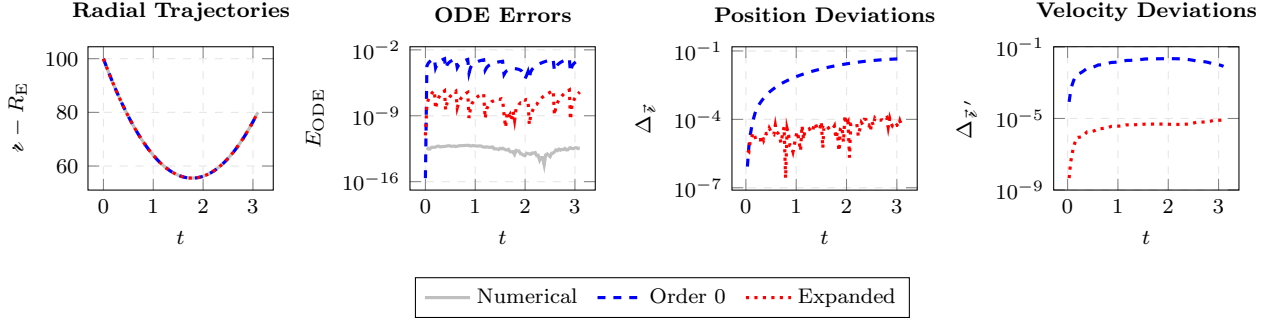
\addMatlabQuadFigureSmall{lpd__data__raindrop.csv}{Data for a raindrop (generated with inputs from Table \ref{tab:fig_inputs}).  Note that this trajectory achieves a minimum radius above the surface of the Earth (without being interrupted by $R\f$), providing an example of trajectory Type 9 of Figure \ref{fig:trajectory_types}.}{fig:raindrop}

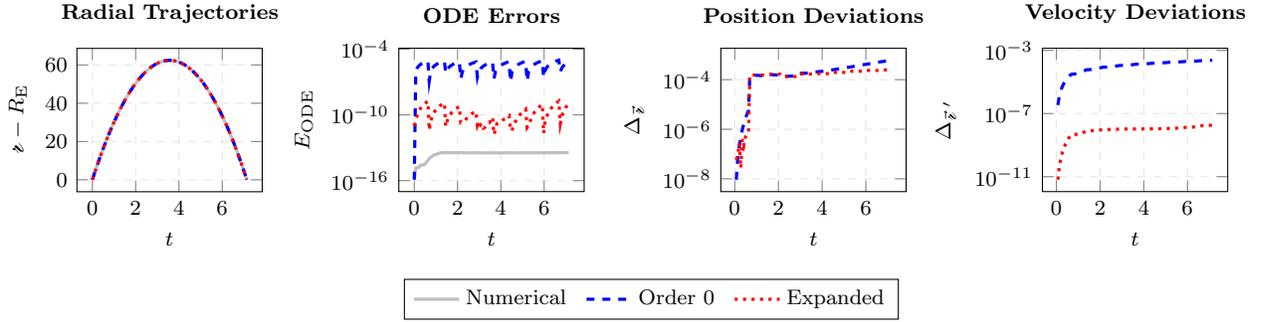
\addMatlabQuadFigureSmall{lpd__data__golf_ball.csv}{Data for a golf ball (generated with inputs from Table \ref{tab:fig_inputs}).  This trajectory is an example of trajectory Type 5 of Figure \ref{fig:trajectory_types}.}{fig:golf_ball}

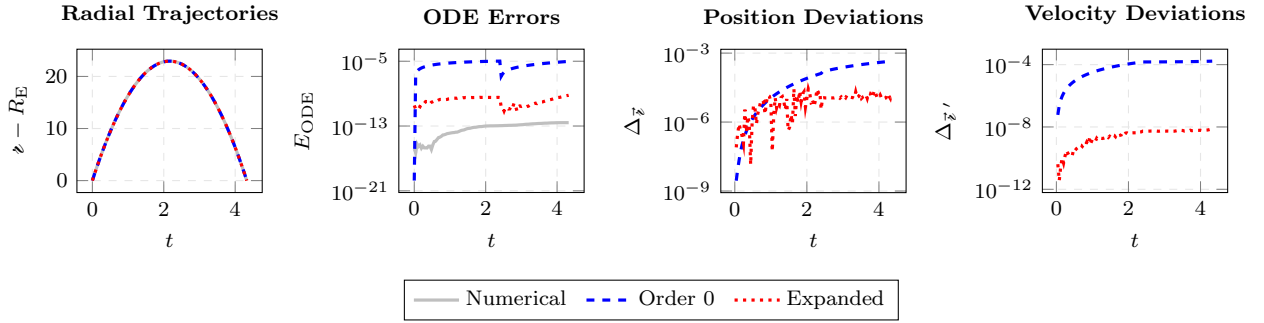
\addMatlabQuadFigureSmall{lpd__data__baseball.csv}{Data for a baseball (generated with inputs from Table \ref{tab:fig_inputs}).  Note that, for $0 < t \lessapprox 0.5$, we have that $( E\subODE )\subtext{Numerical} \ll ( E\subODE )\subtext{Expanded} \ll ( E\subODE )\subtext{Order 0}$ and yet $( \Delta_{ \absposv } )\subtext{Order 0} \ll ( \Delta_{ \absposv } )\subtext{Expanded}$.  In other words, the Expanded and Numerical positions most closely satisfy the ODE \eqref{original_model} but are not closest to one another, pair-wise.  This trajectory is an example of trajectory Type 5 of Figure \ref{fig:trajectory_types}.}{fig:baseball}

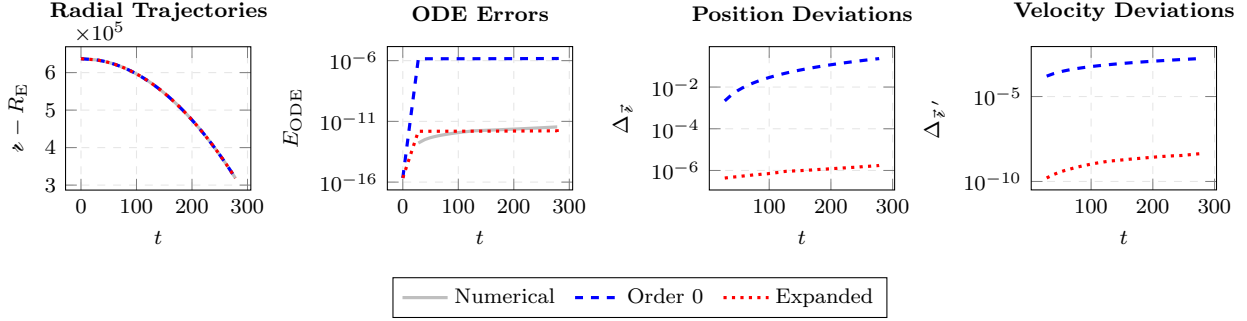
\addMatlabQuadFigureSmall{lpd__data__beach_ball.csv}{Data for a beach ball (generated with inputs from Table \ref{tab:fig_inputs}).  This trajectory is an example of trajectory Type 1 of Figure \ref{fig:trajectory_types}.}{fig:beach_ball}

\subsection{Some observations}\label{sec:observations}

We make the following observations of Figures \ref{fig:point_mass} -- \ref{fig:beach_ball}.
\begin{enumerate}[label={(\textbf{\arabic*})}]

	\item In each figure, we observe that $( E\subODE )\subtext{Expanded} \ll ( E\subODE )\subtext{Order 0}$, providing numerical evidence that our $\O( \eps )$ solutions improve upon our $\O( 1 )$ solutions, in the sense that they reduce the error in the original ODE \eqref{original_model}.
	
	\item In Figure \ref{fig:point_mass}, we see that $( E\subODE )\subtext{Expanded} \ll ( E\subODE )\subtext{Numerical}$, providing numerical evidence that our expanded solutions can be highly accurate with respect to the original ODE \eqref{original_model}, depending on the particular problem and on the chosen values of the error-control parameters.

\end{enumerate}

\section{Conclusion}\label{sec:conclusion}

In this paper, we provided a full 3D treatment of the classic linear projectile problem, subject to several generalizations.  By applying a perturbation technique with small parameter $\eps$, we provided exact integral solutions to $\O( \eps )$ for the general problem and exact closed-form solutions to $\O( \eps )$ for the special case of constant wind.  We investigated the time of tangency, times of flight, and extreme values of the radius achieved by the projectile.  We developed a method to control the error in our approximations and provided algorithms utilizing this method.  We then provided numerical evidence that our $\O( \eps )$ solutions increase the accuracy of our $\O( 1 )$ solutions with respect to the ODE modeling the physical problem.  We have also made freely available our Matlab code that reproduces all the numerical data presented in this paper.

\appendix

\section{Selected calculations of the integrals $I_0$ and $I_1$}\label{sec:I_n}

We find, for $b\eff \neq 0$, that
\begin{subequations}% See pg 7 of Projectile_2021_04_25.
\begin{align}
\I{ 0 }{ x \mapsto 1 }( \tau )
	& = \int_0^\tau e^{ \tau' / \td } \, d\tau'\\
	& = \td ( \exppt - 1 ),\\
\expmt \I{ 0 }{ x \mapsto 1 }( \tau )
	& = \td ( 1 - \expmt ),\\
\I{ 1 }{ x \mapsto 1 }( \tau )
	& = \int_0^\tau e^{ -\tau' / \td } \I{ 0 }{ x \mapsto 1 }( \tau' ) \, d\tau'\\
	& = \td \int_0^\tau ( 1 - e^{ -\tau' / \td } ) \, d\tau'\\
	& = \td^2 ( \expmt - 1 ) + \td \tau,
\end{align}
\end{subequations}
\begin{subequations}
\begin{align}
\I{ 0 }{ x \mapsto x }( \tau )
	& = \int_0^\tau e^{ \tau' / \td } \tau' \, d\tau'\\
	& = \td^2 ( 1 - \exppt ) + \td \tau \exppt,\\
\expmt \I{ 0 }{ x \mapsto x }( \tau )
	& = \td^2 ( \expmt - 1 ) + \td \tau,\\
\I{ 1 }{ x \mapsto x }( \tau )
	& = \int_0^\tau e^{ -\tau' / \td } \I{ 0 }{ x \mapsto x }( \tau' ) \, d\tau'\\
	& = \int_0^\tau \td^2 ( e^{ -\tau' / \td } - 1 ) + \td \tau' \, d\tau'\\
	& = \td \tau ( \tfrac{ 1 }{ 2 } \tau - \td ) + \td^3 ( 1 - \expmt ),
\end{align}
\end{subequations}
\begin{subequations}
\begin{align}
\I{ 0 }{ x \mapsto e^{ -x / \td } }( \tau )
	& = \int_0^\tau \, d\tau'\\
	& = \tau,\\
\expmt \I{ 0 }{ x \mapsto e^{ -x / \td } }( \tau )
	& = \tau \expmt,\\
\I{ 1 }{ x \mapsto e^{ -x / \td } }( \tau )
	& = \int_0^\tau e^{ -\tau' / \td } \I{ 0 }{ x \mapsto e^{ -x / \td } }( \tau' ) \, d\tau'\\
	& = \int_0^\tau e^{ -\tau' / \td } \tau' \, d\tau'\\
	& = \td^2 ( 1 - \expmt ) - \td \tau \expmt,
\end{align}
\end{subequations}
\begin{subequations}
\begin{align}
\I{ 0 }{ x \mapsto x e^{ -x / \td } }( \tau )
	& = \int_0^\tau \tau' \, d\tau'\\
	& = \tfrac{ 1 }{ 2 } \tau^2,\\
\expmt \I{ 0 }{ x \mapsto x e^{ -x / \td } }( \tau )
	& = \tfrac{ 1 }{ 2 } \tau^2 \expmt,\\
\I{ 1 }{ x \mapsto x e^{ -x / \td } }( \tau )
	& = \int_0^\tau e^{ -\tau' / \td } \I{ 0 }{ x \mapsto x e^{ -x / \td } }( \tau' ) \, d\tau'\\
	& = \frac{ 1 }{ 2 } \int_0^\tau e^{ -\tau' / \td } ( \tau' )^2 \, d\tau'\\
	& = \td^3 ( 1 - \expmt ) - \expmt \td \tau ( \td + \tfrac{ 1 }{ 2 } \tau ),
\end{align}
\end{subequations}
\begin{subequations}
\begin{align}
\I{ 0 }{ x \mapsto e^{ -2 x / \td } }( \tau )
	& = \int_0^\tau e^{ -\tau' / \td } \, d\tau'\\
	& = \td ( 1 - \expmt ),\\
\expmt \I{ 0 }{ x \mapsto e^{ -2 x / \td } }( \tau )
	& = \td \expmt ( 1 - \expmt ),\\
\I{ 1 }{ x \mapsto e^{ -2 x / \td } }( \tau )
	& = \int_0^\tau e^{ -\tau' / \td } \I{ 0 }{ x \mapsto e^{ -2 x / \td } }( \tau' ) \, d\tau'\\
	& = \td \int_0^\tau e^{ -\tau' / \td } ( 1 - e^{ -\tau' / \td } ) \, d\tau'\\
	& = \tfrac{ 1 }{ 2 } \td^2 ( \expmt - 1 )^2.
\end{align}
\end{subequations}

\section{Expansions in orders of $b\eff$}\label{sec:b_eff_expansions}

Here we effectively apply a perturbation technique using the quantity $b\eff$, which we assume to be small.  More specifically, we assume that the condition
\bq
\label{bs}
b\eff < b\eff^\star\tag{\textbf{bs}}
\eq
holds for some appropriately chosen\footnote{See Section \ref{sec:traj_plots} for an example value of $b\eff^\star$.} parameter $b\eff^\star \ll 1$.

\subsection{Expanded positions and velocities}\label{sec:b_eff_expansion_solns}

We write the Ansatz
\bq
\label{b_eff_expansion_ansatz}
\rhov_0( \tau ; \bs , \cw ) = \rhovsup{ 0 }_0( \tau ; \bs , \cw ) + b\eff \rhovsup{ 1 }_0( \tau ; \bs , \cw ) + \O( b\eff^2 ),
\eq
where the components $\rhovsup{ j }_0$ are to be determined.  To find these components, we can equivalently (\textbf{1}) Taylor-expand the general solution $\rhov_0( \tau ; \bnz , \cw )$ in powers of $b\eff$ or (\textbf{2}) re-cast the IVP for $\rhov_0( \tau ; \bnz , \cw )$ into one IVP for each order of $b\eff$ in which we're interested.

We find that
\begin{subequations}
\begin{align}
\rhovsup{ 0 }_0( \tau ; \bs , \cw )      & = \rhov_0( \tau ; \bz , \cw ),\\
( \rhovsup{ 0 }_0 )'( \tau ; \bs , \cw ) & = \rhovprime_0( \tau ; \bz , \cw ),\\
\rhovsup{ 1 }_0( \tau ; \bs , \cw )      & = \tfrac{ 1 }{ 2 } ( \w_0 - \dv ) \tau^2 - \tfrac{ 1 }{ 6 } \gv \tau^3,\\
( \rhovsup{ 1 }_0 )'( \tau ; \bs , \cw ) & = ( \w_0 - \dv ) \tau - \tfrac{ 1 }{ 2 } \gv \tau^2.
\end{align}
\end{subequations}
The zeroth-order solutions to the original problem \eqref{original_model} are defined in the usual way, for $0 \le t \le t\c \tau_\star$:
\begin{subequations}
\begin{empheq}[box=\mybox]{align}
\absposv\subtext{order 0}( t ; \bs , \cw ) 		& \equiv \absposv_0 + r\c \rhov_0\left( \tau = \frac{ t }{ t\c } ; \bs , \cw \right),\\
\absposvprime\subtext{order 0}( t ; \bs , \cw )	& \equiv v\c \rhovprime_0\left( \tau = \frac{ t }{ t\c } ; \bs , \cw \right).
\end{empheq}
\end{subequations}
Applying the same technique, we find the components of $\rhov_1( \tau ; \bs , \cw )$ to be given by
\begin{subequations}
\begin{align}
\rhovsup{ 0 }_1( \tau ; \bs , \cw )      & = \rhov_1( \tau ; \bz , \cw ),\\
( \rhovsup{ 0 }_1 )'( \tau ; \bs , \cw ) & = \rhovprime_1( \tau ; \bz , \cw ),\\
\rhovsup{ 1 }_1( \tau ; \bs , \cw )      & = -2 \p_1 ( e^{ -\tau } - 1 ) - 2 \p_1 \tau + \p_1 \tau^2 - \tfrac{ 1 }{ 3 } \p_1 \tau^3 + ( \tfrac{ 1 }{ 4 } \p_3 - \p_4 ) \tau^4 + \tfrac{ 1 }{ 5 } \p_4 \tau^5,\\
( \rhovsup{ 1 }_1 )'( \tau ; \bs , \cw ) & = 2 \p_1 ( e^{ -\tau } - 1 ) + 2 \p_1 \tau - \p_1 \tau^2 + 4 ( \tfrac{ 1 }{ 4 } \p_3 - \p_4 ) \tau^3 + \p_4 \tau^4,
\end{align}
\end{subequations}
where
\begin{subequations}
\begin{align}
\p_2 & \equiv -\tfrac{ 1 }{ 2 } \g ( \dv - \w_0 ) + \tfrac{ 3 }{ 2 } \gv ( \d \s - \av \cdot \w_0 ) + \tfrac{ 1 }{ 2 } ( 3 \d \s - \av \cdot \w_0 ) ( \dv - \w_0 ) b\eff',\\
\p_3 & \equiv \tfrac{ 1 }{ 3 } \g \gv + \tfrac{ 2 }{ 3 } b\eff' \g ( \dv - \w_0 ) + b\eff' \gv ( \d \s - \tfrac{ 1 }{ 2 } \av \cdot \w_0 ),\\
\p_4 & \equiv \tfrac{ 5 }{ 12 } \g \gv b\eff',\\
\p_1 & \equiv -\p_2 + 3 \p_3 - 12 \p_4.
\end{align}
\end{subequations}
Neglecting terms of order $b\eff^2$ and higher, we then write the (doubly) expanded dimensionless position, for $0 \le \tau \le \tau_\star$, as
\begin{subequations}
\begin{empheq}[box=\mybox]{align}
\rhov\comp( \tau ; \bs , \cw )	& \equiv \rhov_0( \tau ; \bs , \cw ) + \eps \rhov_1( \tau ; \bs , \cw )\\
								& = \left[ \rhovsup{ 0 }_0( \tau ; \bs , \cw ) + b\eff \rhovsup{ 1 }_0( \tau ; \bs , \cw ) \right]\\
								\nonumber%
								& \quad + \eps \left[ \rhovsup{ 0 }_1( \tau ; \bs , \cw ) + b\eff \rhovsup{ 1 }_1( \tau ; \bs , \cw ) \right],
\end{empheq}
\end{subequations}
with the (doubly) expanded dimensionless velocity $\rhovprime\comp$ defined similarly.  The (doubly) expanded position $\absposv\comp$ and velocity $\absposvprime\comp$ are then defined accordingly, for $0 \le t \le t\c \tau_\star$:
\begin{subequations}
\label{expanded_soln_bs}
\begin{empheq}[box=\mybox]{align}
\absposv\comp( t ; \bs , \cw ) 		& \equiv \absposv_0 + r\c \rhov\comp\left( \tau = \frac{ t }{ t\c } ; \bs , \cw \right),\\
\absposvprime\comp( t ; \bs , \cw )	& \equiv v\c \rhovprime\comp\left( \tau = \frac{ t }{ t\c } ; \bs , \cw \right).
\end{empheq}
\end{subequations}

\subsection{Expanded times of tangency and flight}

Using Ansatzes of the same form as \eqref{b_eff_expansion_ansatz}, we find that
\begin{subequations}
\label{tau_bs}
\begin{align}
( \tau\tang[ \bs , \cw ] )^{ ( 0 ) }			 & = \tau\tang[ \bz ],\\
( \tau\tang[ \bs , \cw ] )^{ ( 1 ) }			 & = \tfrac{ 2 }{ \g } ( \av \cdot \w_0 - \d \s ),\\
( \tau\flight\posttan[ \bs , \cw ] )^{ ( 0 ) } & = \tau\flight\posttan[ \bz , \cw ],\\
( \tau\flight\posttan[ \bs , \cw ] )^{ ( 1 ) } & = \tfrac{ 3 }{ \g } ( \av \cdot \w_0 - \d \s ),\\
( \tau\flight\pretan[ \bs , \cw ] )^{ ( 0 ) }  & = \tau\flight\pretan[ \bz , \cw ],\\
( \tau\flight\pretan[ \bs , \cw ] )^{ ( 1 ) }  & = \tfrac{ 3 }{ \g } ( \av \cdot \w_0 - \d \s ).
\end{align}
\end{subequations}

\end{document}